\RequirePackage[hyphens]{url} 
\documentclass[%
 reprint,
 superscriptaddress,
 showpacs,amsmath,amssymb,
 aps,
floatfix,
nofootinbib
]{revtex4-1}
\usepackage{url}
\usepackage[breaklinks]{hyperref}
\PassOptionsToPackage{hyphens}{url}\usepackage{hyperref}
\usepackage{mathpazo}
\usepackage{times}
\usepackage{graphicx}

\usepackage{dcolumn}%
\usepackage{footnote}
\usepackage{makecell,multirow}
\usepackage{booktabs}
\usepackage[utf8]{inputenc}
\usepackage{tikz}
\usetikzlibrary{shapes,arrows}
\usepackage{multirow}
\usepackage{siunitx}
\usepackage{amsmath}
\usepackage{bm}
\usepackage[mathlines]{lineno}
\usepackage{braket}
\usepackage{color,soul}
\usepackage{subfigure}
\usetikzlibrary{shapes.geometric}
\usepackage{changepage}
\usepackage[bottom]{footmisc}

\usepackage[american,siunitx]{circuitikz}
\usetikzlibrary{arrows,calc,positioning}

\newsavebox{\foobox}

\tikzset{ar/.style={-latex,shorten >=-1pt, shorten <=-1pt}}
\begin{document}
\preprint{APS/123-QED}

\title{Axion Dark Matter eXperiment: Run 1B Analysis Details}
\author{C.Bartram}\email[Correspondence to: ]{chelsb89@uw.edu}
\affiliation{University of Washington, Seattle, WA 98195, USA}
\author{T. Braine}
  \affiliation{University of Washington, Seattle, WA 98195, USA}
\author{R. Cervantes}
  \affiliation{University of Washington, Seattle, WA 98195, USA}
 \author{N. Crisosto}
   \affiliation{University of Washington, Seattle, WA 98195, USA}
\author{N. Du}%
  \affiliation{University of Washington, Seattle, WA 98195, USA}
\author{G. Leum}
  \affiliation{University of Washington, Seattle, WA 98195, USA}
 \author{L. J Rosenberg}%
  \affiliation{University of Washington, Seattle, WA 98195, USA}
  \author{G. Rybka}%
  \affiliation{University of Washington, Seattle, WA 98195, USA}
    \author{J.~Yang}%
  \affiliation{University of Washington, Seattle, WA 98195, USA}
  
%
\author{D. Bowring}
  \affiliation{Fermi National Accelerator Laboratory, Batavia IL 60510, USA}
\author{A. S. Chou} 
  \affiliation{Fermi National Accelerator Laboratory, Batavia IL 60510, USA}
    \author{R. Khatiwada}
     \affiliation{Fermi National Accelerator Laboratory, Batavia IL 60510, USA}
  \affiliation{Illinois Institute of Technology, Chicago IL 60616, USA}
\author{A. Sonnenschein} 
  \affiliation{Fermi National Accelerator Laboratory, Batavia IL 60510, USA}
  \author{W. Wester} 
  \affiliation{Fermi National Accelerator Laboratory, Batavia IL 60510, USA}

\author{G. Carosi}
\affiliation{Lawrence Livermore National Laboratory, Livermore, CA 94550, USA}
\author{N. Woollett}
\affiliation{Lawrence Livermore National Laboratory, Livermore, CA 94550, USA}

\author{L. D. Duffy}
  \affiliation{Los Alamos National Laboratory, Los Alamos, NM 87545, USA}

\author{M. Goryachev}
\affiliation{University of Western Australia, WA, Australia}
\author{B. McAllister}
\affiliation{University of Western Australia, WA, Australia}
\author{M. E. Tobar}
\affiliation{University of Western Australia, WA, Australia}

\author{C. Boutan}
  \affiliation{Pacific Northwest National Laboratory, Richland, WA 99354, USA}
\author{M. Jones}
  \affiliation{Pacific Northwest National Laboratory, Richland, WA 99354, USA}
\author{B. H. LaRoque}
  \affiliation{Pacific Northwest National Laboratory, Richland, WA 99354, USA}
\author{N. S.~Oblath}
  \affiliation{Pacific Northwest National Laboratory, Richland, WA 99354, USA}
\author{M.~S. Taubman}
  \affiliation{Pacific Northwest National Laboratory, Richland, WA 99354, USA}

\author{John Clarke}
  \affiliation{University of California, Berkeley, CA 94720, USA}
\author{A. Dove}
  \affiliation{University of California, Berkeley, CA 94720, USA}
\author{A. Eddins}
  \affiliation{University of California, Berkeley, CA 94720, USA}
\author{S. R. O'Kelley}
  \affiliation{University of California, Berkeley, CA 94720, USA}
\author{S. Nawaz}
  \affiliation{University of California, Berkeley, CA 94720, USA}
\author{I. Siddiqi}
  \affiliation{University of California, Berkeley, CA 94720, USA}
\author{N. Stevenson}
  \affiliation{University of California, Berkeley, CA 94720, USA}

\author{A. Agrawal}
\affiliation{University of Chicago, IL 60637, USA}
\author{A. V. Dixit}
\affiliation{University of Chicago, IL 60637, USA}
 
\author{J.~R.~Gleason}
  \affiliation{University of Florida, Gainesville, FL 32611, USA}
\author{S. Jois}
  \affiliation{University of Florida, Gainesville, FL 32611, USA}
 \author{P. Sikivie}
  \affiliation{University of Florida, Gainesville, FL 32611, USA}
  \author{J. A. Solomon}
  \affiliation{University of Florida, Gainesville, FL 32611, USA}
\author{N. S. Sullivan}
  \affiliation{University of Florida, Gainesville, FL 32611, USA}
\author{D. B. Tanner}
  \affiliation{University of Florida, Gainesville, FL 32611, USA}

\author{E.~Lentz}
  \affiliation{University of G\"{o}ttingen, G\"{o}ttingen, Germany}  
  
\author{E. J. Daw}
  \affiliation{University of Sheffield, Sheffield, UK}
  
  \author{M. G. Perry}
  \affiliation{University of Sheffield, Sheffield, UK}
  
\author{J. H. Buckley}
  \affiliation{Washington University, St. Louis, MO 63130, USA}
\author{P. M. Harrington}
  \affiliation{Washington University, St. Louis, MO 63130, USA}
\author{E. A. Henriksen}
  \affiliation{Washington University, St. Louis, MO 63130, USA} 
\author{K. W. Murch}
  \affiliation{Washington University, St. Louis, MO 63130, USA}
 
\collaboration{ADMX Collaboration}\noaffiliation


\date{\today}
\begin{abstract}
Searching for axion dark matter, the ADMX collaboration acquired data from January to October 2018, over the mass range 2.81--3.31 $\si\micro$eV, corresponding to the frequency range 680--790 MHz. Using an axion haloscope consisting of a microwave cavity in a strong magnetic field, the ADMX experiment excluded Dine-Fischler-Srednicki-Zhitnisky (DFSZ) axions at 100\% dark matter density over this entire frequency range, except for a few gaps due to mode crossings. This paper explains the full ADMX analysis for Run 1B, motivating analysis choices informed by details specific to this run. 
\end{abstract}

\maketitle
\section{Introduction}
An abundance of astrophysical observations indicate that the majority (85\%~\cite{Planck}) of the mass of the universe exists in some unidentified form, called `dark matter'. The Lambda cold dark matter ($\Lambda$CDM) model of the universe ascribes the following characteristics to the dark matter: that it is feebly interacting, non-relativistic, and non-baryonic~\cite{PDG}. One dark matter candidate, known as the axion, solves the so-called strong CP (Charge-Parity) problem via a global chiral symmetry introduced by Peccei and Quinn~\cite{Peccei1977June,Weinberg:1977ma,Wilczek:1977pj}. Assuming a typical post-inflationary scenario, QCD (Quantum Chromodynamics) axions in a mass range of 1--100 $\si\micro$eV may account for the entirety of dark matter, if they exist~\cite{PhysRevD.96.095001,PhysRevLett.118.071802,Borsanyi2016}. Two models, the KSVZ (Kim-Shifman-Vainshtein-Zakharov) model~\cite{Kim:1979if,Shifman:1979if} and the DFSZ (Dine-Fischler-Srednicki-Zhitnisky) model~\cite{Dine:1981rt,Zhitnitsky:1980tq}, are benchmarks for axion experiments and can be described by their coupling strengths of the axion to photons. The dimensionless axion-photon coupling parameter, known as $g_{\gamma}$, is smaller for DFSZ axions than KSVZ axions by a factor of approximately 2.7, making DFSZ axions more challenging to detect. In both models, the strength of the axion coupling to photons is further suppressed by the very high energy scale associated with the Peccei-Quinn (PQ) symmetry breaking. The dimensionless coupling, $g_{\gamma}$ is related to the axion coupling to two photons via $g_{a\gamma\gamma}={\alpha}g_{\gamma}/{{\pi}f_a}$, where $\alpha$ is the fine structure constant, and $f_a$ is the PQ symmetry breaking scale. The DFSZ axion couples directly to both hadrons and leptons, whereas the KSVZ axion couples directly only to hadrons. In all grand unified theories, the coupling strength of the axion to two photons is that of the DFSZ model~\cite{PhysRevLett.47.402}.

Although a number of experimental efforts to detect axions are now underway, the Sikivie microwave cavity detector~\cite{Sikivie:1983ip,PhysRevLett.52.695.2}, marked the first feasible means of detecting the so-called `invisible' axion. This paper described the first axion haloscope, in which a static magnetic field provided a new channel for the axion to decay into a photon. The process, known as inverse Primakoff conversion~\cite{PhysRevD.84.121302}, follows from the equations of axion electrodynamics. The resulting excess power from the photon could then be resonantly enhanced and detected in a microwave cavity. A few years ago, the Axion Dark Matter eXperiment, ADMX, became the first experiment to reach DFSZ sensitivity. Defined as `Run 1A', this run resulted in the reporting of a limit on $g_{a\gamma\gamma}$ over axion masses of 2.68--2.7 $\si\micro$eV~\cite{Du_2018}. The experiment recently extended this limit to cover the range from 2.81--3.31 $\si\micro$eV, corresponding to a  frequency range from about 680 to 790 MHz. The resulting data, acquired over a period between January and October of 2018, are referred to as `Run 1B'~\cite{PhysRevLett.124.101303}. This paper gives complete details of the analysis for Run 1B,  
assuming a fully virialized dark matter halo. While the foundation of the analysis is unchanged from previous runs, improvements have been made, and the details specific to this run are explained. 

There are two key components to a haloscope analysis worth emphasizing: axion search data and noise characterization data. The former is acquired by digitizing power from the cavity, in series with a number of other processes (described as the `run cadence'), whereas the latter is acquired periodically by halting axion search operations and performing a noise temperature measurement. Both are essential to the final analysis. 

Ultimately, the analysis hinges not only on these two distinct sets of data, but on a number of other factors, which are described in the course of this paper, and outlined below.
\begin{enumerate}
\item The experimental configuration is described for Run 1B (Section~\ref{sec:experiment}), with particular emphasis on the aspects of the receiver chain that were updated for this run. For the purposes of this paper, the receiver chain is defined as all RF components that are used in both axion search and noise characterization modes, as described in Section~\ref{sec:experiment}. The design of the receiver chain directly motivates particular choices for the analysis. 
\item Section~\ref{sec:data_taking} undertakes a discussion of the run cadence and means of data acquisition. This section includes the acquisition of sensor data as well as radio frequency (RF) data. The specifics of the data pre-processing are elaborated.
\item The techniques that were used to characterize the system noise temperature, which is critical to quantifying our sensitivity, are explained in Section~\ref{section:noise_measurement}. This section also enumerates and motivates data quality cuts. Systematic uncertainties are quantified and discussed. 
\item Section~\ref{sec:analysis_procedure} explains the analysis of the raw power spectra, beginning with removal of the warm electronics baseline, followed by the filtering and combining of data to form the \emph{grand spectrum} via an optimal weighting procedure. 
\item Section~\ref{sec:synthetics} describes both hardware and software synthetic axion injections.
\item Section~\ref{sec:mode_crossings} describes the handling of mode crossings. 
\item Section~\ref{sec:rescan_procedure} explains the rescan procedure.
\item The final section of this paper (Section~\ref{sec:limit}) explains the limit-setting procedure and interpretation.\\
\end{enumerate}
\noindent Barring the existence of any persistent candidates, the limit setting process marks the final step in the data-processing sequence, resulting in a statement of exclusion over the Run 1B frequency range.
\vspace{0.5cm}
\section{Experimental Setup}
\label{sec:experiment}
\subsection{Detector}

The Axion Dark Matter eXperiment uses the haloscope approach to search for dark matter axions~\cite{Sikivie1985,Sikivie:1983ip}. A cavity haloscope is a high-$Q$, cryogenic, microwave cavity immersed in a high field solenoid. The ADMX solenoid can be operated at fields as high as 8.5 T, but, in the interest of safety and reliability, was operated at 7.6 T throughout the course of Run 1B. The Run 1B cavity consisted of a 140-liter cavity made of copper-plated stainless steel (136 liter when the tuning rod volume is subtracted). Two 50.8-mm diameter copper tuning rods ran the length of the cavity parallel to the axis. Each rod could be translated from near the wall to near the center of the cavity. To detect the axion signal, the microwave cavity must be tuned to match the signal frequency defined by $f_a\,{\approx}\,m_a$ (not accounting for its small kinetic energy). The axion mass is unknown over a broad range, so the cavity was tuned by moving metallic rods to scan a range of frequencies. Power from the cavity was extracted by an antenna consisting of the exposed center conductor of a semi-rigid coaxial cable. The antenna was inserted into the top of the cavity and connected to the receiver chain. Assuming their existence, axions would deposit excess power in the cavity when the cavity was tuned to the axion mass equivalent frequency. This excess power would be detected as a small narrowband excess in the digitized spectrum. The detected axion power is given by
\begin{widetext}
\begin{multline}
    P_{\text{axion}}=2.2{\times}10^{-23}\mathrm{W}\left(\frac{\beta}{1+\beta}\right)\left(\frac{V}{136~\mathrm{\ell}}\right)\left(\frac{B}{7.6~ \mathrm{T}}\right)^2\left(\frac{C_{010}}{0.4}\right)\left(\frac{g_{\gamma}}{0.36}\right)^2\left(\frac{\rho}{0.45~\mathrm{GeV cm^{-3}}}\right) \\ \left(\frac{Q_{\text{axion}}}{10^6}\right)\left(\frac{f}{740~ \mathrm{MHz}}\right)\left(\frac{Q_{\text{L}}}{30{,}000}\right)\left(\frac{1}{1+(2{\delta}f_{a}/{{\Delta}f})^2}\right),
\label{eqn:axion_pwr}
\end{multline}
\end{widetext}
\noindent where $V$ is the volume of the cavity, $B$ is the static magnetic field from the solenoid, $\rho$ is the dark matter density, $f$ is the frequency of the photon, $Q_{\text{L}}$ is the loaded quality factor, $Q_{\text{axion}}$ is the axion quality factor, and $C_{010}$ is the form factor. The form factor describes the overlap of the electric field of the cavity mode and magnetic field generated by the solenoid~\cite{Sikivie1985}. The indices denote the usage of the $\mathrm{TM_{010}}$ mode, which maximizes the form factor. The cavity mode linewidth is given by ${\Delta}f=f/Q_{\text{L}}$. The detuning factor, ${\delta}f_{a}$, is some frequency offset from the cavity resonance. The cavity coupling parameter, which describes how much power is picked up by the strongly coupled antenna, is given by $\beta=\left(Q_0/Q_{\text{L}}-1\right)$, where $Q_0$ is the unloaded cavity quality factor. The dark matter density of 0.45 $\mathrm{GeV/cm^{3}}$~\cite{Read_2014} has previously been assumed by ADMX in presenting its sensitivity. Of note is that the deposited power is on the order of 10s of yoctowatts--a level which is just barely detectable using state-of-the-art technology. Typically, the experimentalist has control over the cavity coupling parameter, volume, magnetic field, form factor and quality factor, whereas the remaining parameters are set by nature. Optimizing for signal-to-noise (SNR) means maximizing the former, while minimizing the system noise.

ADMX Run 1B relied on two key components to achieve DFSZ sensitivity: the use of a quantum amplifier, and a dilution refrigerator. The quantum amplifier afforded the experiment a low amplifier noise, whereas the dilution refrigerator reduced the physical temperature of the microwave cavity and the quantum amplifier. Combined, the two advances reduced the system noise compared to earlier ADMX experiments~\cite{PhysRevLett.80.2043,PhysRevD.69.011101,Asztalos:2009yp}.

ADMX has evolved and been improved since its first run at DFSZ sensitivity~\cite{PhysRevLett.120.151301}. Each run presents its own unique set of challenges, motivating unique choices for the analysis. Challenges pertaining to the Run 1B receiver chain will be described in the following sections.

\subsection{ADMX Run 1B Receiver Chain}

\begin{figure}
    \centering
    \includegraphics[width=0.4\textwidth]{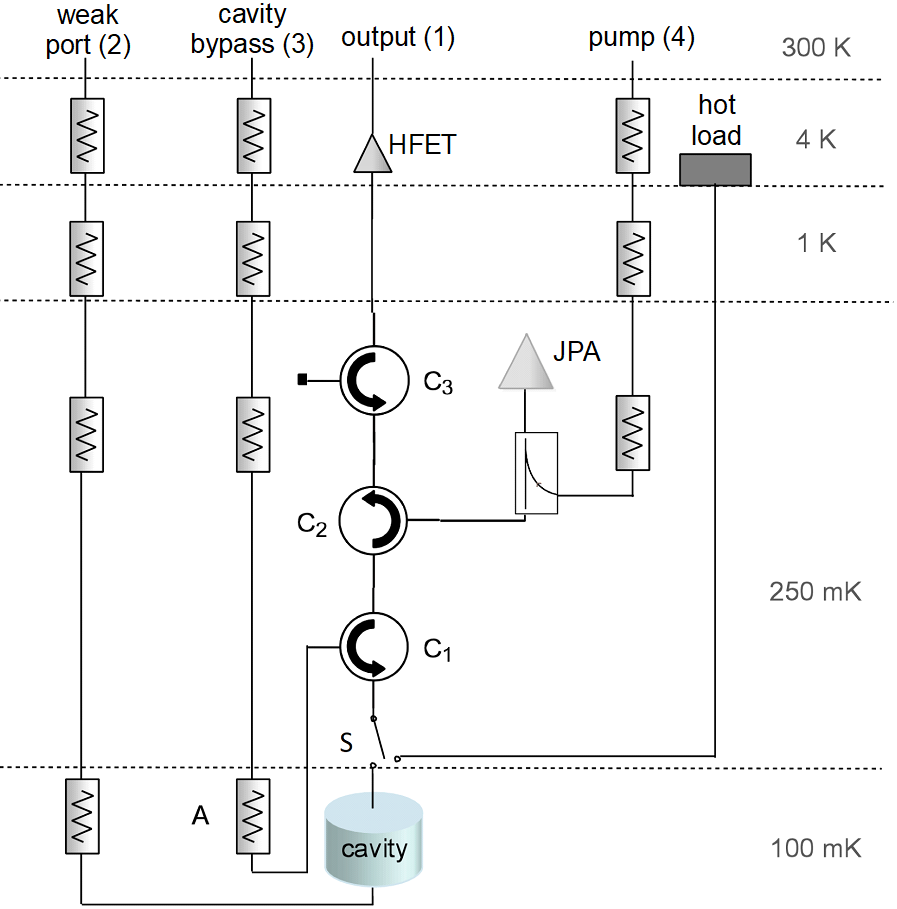}
   \caption{ADMX Run 1B receiver chain. $C_{1}$, $C_{2}$ and $C_{3}$ are circulators. The temperature stages for all components are shown on the right-hand side. Attenuator labelled $\mathrm{A}$ plays an important role in noise calibration.}
    \label{fig:run1b_receiver_chain}
\end{figure}

The receiver chain for ADMX varies between runs, as the system is continuously optimized for the frequency range covered. For Run 1B, the part of the receiver chain that was contained in the cold space (defined as everything that is colder than room temperature) is shown in Fig.~\ref{fig:run1b_receiver_chain}. The receiver chain was designed with two goals in mind: first, to read out power from the cavity (`axion search mode') and second, to characterize the noise of the receiver chain (`noise characterization mode'). There were a few factors which motivated the design of the operating modes, each accessible by flipping an RF switch (indicated by $\mathrm{S}$ in Fig.~\ref{fig:run1b_receiver_chain}) that allowed the JPA to be connected to either the cavity (axion search mode) or the hot load (noise characterization mode). The design of the axion search mode was driven by the desire to minimize attenuation along the output line and reduce the amplifier and physical noise as much as possible. Likewise, the design of the noise characterization mode was motivated by the need to have a reliable means of heating the 50-ohm terminator (`hot load') at the end of the output line, as described in Section~\ref{section:noise_measurement}. 
With the switch configured to connect the output line to the cavity, there were three critical RF paths. First, a swept RF signal from the vector network analyzer (VNA) could be routed through the cavity via the weak port (2) and up through the cavity and output line (1), back to the VNA. The weak port is aptly named to describe the fact that it connects to a weakly coupled antenna at the base of the cavity. Such measurements were referred to as \emph{transmission measurements}. Next, a swept RF signal could be injected via the bypass line (3), reflected off the cavity and emerge via the output line (1). Because this setup was used to measure power reflected off the cavity, this is referred to colloquially as a \emph{reflection measurement}, even though the signal path technically followed that of an $\mathrm{S_{21}}$ measurement. While the axion search data were being acquired, connections to network analyzer input and output were disabled and power coming out of the cavity via the output line (1) was amplified, mixed to an IF frequency, filtered, and further amplified before reaching the digitizer (Signatec PX1500~\cite{Signatec}). The other two setups (reflection and transmission routes) were used to characterize the detector and receiver chain. Reflection measurements were used to determine and adjust the antenna coupling, and transmission measurements were used to determine the cavity quality factor and resonant frequency. Broadly speaking, both measurements were used throughout data-taking operations to check the integrity of the receiver chain, as abnormal transmission or reflection measurements could be indicative of problems along the signal path. 

In the cold space, signals from the cavity on the output line were amplified by a Josephson Parameteric Amplifier (JPA)~\cite{PhysRevB.83.134501,PhysRevApplied.8.054030} followed by a Heterostructure Field-Effect Transistor (HFET), model number LNF-LNC03\_14A from Low Noise Factory~\cite{LNF}. In general, the noise contribution from the first stage amplifier was the dominant source of noise coming from the electronics~\cite{friis1944}, motivating the decision to place the JPA, with its exceedingly low amplifier noise, as close to the strongly coupled antenna as possible. The JPA was highly sensitive to magnetic fields, and was therefore strategically placed in a low-field region, accomplished via a bucking coil that partially cancels the main magnetic field about a meter above the cavity. The JPA was also encased in passive magnetic shielding consisting of a mu-metal cylinder. For the purposes of this paper, all RF electronics from the HFET to the warm electronics are defined as the `downstream' electronics. Further, all components from the first circulator, $C_{1}$, to the third circulator, $C_{3}$, including the JPA, are defined as the `quantum electronics package'. The quantum electronics package was contained within a metal framework that is thermally sunk to the top of the cavity. This package was contained in the 250 mK temperature space shown in Fig.~\ref{fig:run1b_receiver_chain}. 

\begin{figure*}
\includegraphics[width=5in]{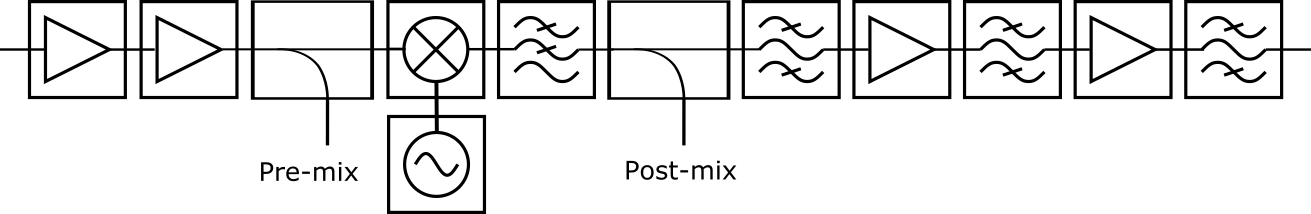}
\caption{Components within the ADMX Run 1B receiver box. From left to right: DC amplifiers (Minicircuits ZX60-3018G+), directional coupler (Minicircuits ZX30-17-5-S+), Polyphase Microwave image-reject mixer, low pass filter (Minicircuits ZX75LP-50+), directional coupler (Minicircuits ZX30-17-5-S+), 2-MHz bandpass filter (Minicircuits SBP-10.7+), DC amplifier (Minicircuits ZFL-500+), 2-MHz bandpass filter (Minicircuits SBP-10.7+), DC amplifier (Minicircuits ZFL-500+), 150-kHz wide custom made filter. The center frequency of the two filters was 10.7 MHz. The intent of these filters is to reduce wide band noise that would cause the digitizer to clip. The directional couplers enable trouble-shooting before and after the mixing stage.}
\label{fig:receiver_components}
\end{figure*}

Upon exiting the insert, signals on the output line entered the warm electronics. First, the signal was amplified by a post-amplifier located immediately outside the insert. The signal then proceeded to the receiver box. The chain of components inside the receiver box can be seen in Fig.~\ref{fig:receiver_components}. The signal from the cavity output was first amplified, then mixed with a local oscillator, before being filtered via a low pass filter, amplified and further filtered, first by a 2-MHz bandpass filter, and later by a 150-kHz bandpass filter.
Upon exiting the receiver box, the signal was digitized with a Nyquist sampling time of 10 ms, yielding a 48.8-kHz wide spectrum centered at the cavity frequency with bins 95-Hz wide. The native digitizer sampling rate itself was 200 Megasamples per second, which was downsampled to 25 Megasamples per second. For each bin, 10,000 of the 10-ms subspectra were co-added to produce the power spectrum from the cavity averaged over 100 s. The noise in each spectrum bin can be reliably approximated as Gaussian. Further instrumentation details can be found in Ref.~\cite{khatiwada2020axion}. There were two data output paths: one for the medium-resolution analysis (this paper) and another for the high-resolution analysis, which is currently in preparation. For the medium-resolution analysis, the 100 s of data were averaged, resulting in a 512-point power spectrum with 95-Hz bin widths. For the high-resolution analysis, an inverse FFT was performed with sufficient phase coherence to be able to reconstruct the characteristics of the time series. The 100-s digitization time was a prerequisite for performing a high-resolution search~\cite{PhysRevD.94.082001}. The high-resolution analysis would be able to detect annual and diurnal shifts in the frequency of an axion signal if detected, something unresolvable with the medium-resolution.
\section{Run Cadence}
\label{sec:data_taking}
The goal of an axion haloscope analysis is to search for power fluctuations above an average noise background that could constitute an axion signal. Rescans are used to identify persistent candidates and rule out candidates that arise from statistical fluctuations. For an axion signal to trigger a rescan, it must be flagged as a candidate in the analysis. In ADMX Run 1B, there were three distinct types of candidates, which are explained in Section~\ref{sec:rescan_procedure}, but, in general, a candidate can be thought of as a power fluctuation above the average noise background. With this in mind, the raw data were processed in such a way that accounted for variations in the individual spectra both at a single frequency and across a range of frequencies. 

An axion haloscope search must incorporate mechanisms for discerning false signals from a true signal. Possible false signals include statistical fluctuations, RF interference, and intentionally injected synthetic axion signals. For ADMX Run 1B, such false signals were rejected via both data quality cuts as well as the rescan procedure, described in Sections~\ref{sec:analysis_procedure} and \ref{sec:rescan_procedure}.

\begin{figure*}
\begin{adjustwidth}{}{-8em}
\centering
\dimendef\prevdepth=0
\tikzset{
    every node/.style={
        font=\scriptsize
    },
    decision/.style={
        shape=rectangle,
        minimum height=0.5cm,
        text width=2.cm,
        text centered,
        rounded corners=1ex,
        draw,
        label={[rotate=90,xshift=0.2cm,yshift=0.45cm,color=red]left:NRR},
        label={[rotate=-90,yshift=0.5cm,xshift=-0.1cm,color=blue]right:RR},
    },
    outcome/.style={
        shape=rectangle,
        rounded corners=1ex,
        text width=2.cm,
        minimum height=0.5cm,
        fill=gray!15,
        draw,
        text centered
    },
     found/.style={
        shape=circle,
        fill=blue!15,
        draw,
        text width=1.0cm,
        text centered
    },
    decision tree/.style={
        edge from parent path={[-latex] (\tikzparentnode) -| (\tikzchildnode)},
        sibling distance=2.9cm,
        level distance=1.4cm
    }
}

\begin{tikzpicture}

\node [decision] { First pass through nibble at fixed tuning rate.}
    [decision tree]
    child { node [outcome] { Continue to next nibble. } }
    child { node [decision] { Rescan at variable tuning rate. $\sim$2-5x} 
        child { node [outcome] { Continue to next nibble. } }
        child { node [decision] { Persistence check. } 
            child { node [outcome] { Continue to next nibble. } }
        child { node [decision] { Turn off primary synthetic axion injections. Rescan. } 
        child { node [outcome] { Continue to next nibble. } }
        child { node [decision] { Persistence check. } 
            child { node [outcome] { Continue to next nibble. } }
            child { node [decision] { Make RFI checks. } 
        child { node [outcome] { Continue to next nibble. } }
        child { node [decision] { Turn off secondary synthetic axion injections. } 
        child { node [outcome] { Continue to next nibble. } }
        child { node [decision] { Check for signal suppression in $\mathrm{TM_{010}}$ mode. } 
                child { node [outcome] { Continue to next nibble. } }
            child { node [decision] { Check signal $\propto\,\mathrm{B^2}$. } 
                child { node [outcome] { Continue to next nibble. } }
                child { node [found] { Axion found. } } }
        } }}}}}
    };

\node[draw,align=left,rounded corners=1ex] at (0,-10.7) {\textcolor{blue} {RR}: Rescan Regions identified\\ \textcolor{red} {NRR}: No Rescan Regions identified};
\end{tikzpicture}
\caption{Data-taking decision tree. After a first scan through a 10-MHz nibble, the grand spectrum is checked for rescan triggers. If found, further scans are then acquired to assess if any of the rescan triggers are axion candidates. Typically, there are always some rescan triggers on a first pass through the nibble due to the statistics associated with the chosen tuning rate. Non-axionic rescan regions vanish with increasing statistics. Nevertheless, there are usually some rescan regions remaining. If so-called `persistent candidates' still remain, they are evaluated using two tests:  persistence checks and on-off resonance tests. A persistence check verifies that a signal appears in every spectrum (i.e. is not intermittent). An on-off resonance test verifies that the signal maximizes on resonance. Some of these may be intentionally injected synthetic axions. As such, the blind injection team is asked to disable injections, after which, further rescans follow. Should candidates remain, a spectrum analyzer is used to eliminate the possibility that it is an ambient (external) signal, such as a radio station. If the candidate is still viable, the blind injection team is asked to reveal all secondary synthetic injections. If the candidate is not synthetic, a magnet ramp ensues to verify that the signal power is proportional to the magnetic field squared. Candidates that passed this step would be determined as axionic in nature. When no candidates were uncovered at the DFSZ level, a limit was set.}
\label{fig:decisiontree}
\end{adjustwidth}
\end{figure*}
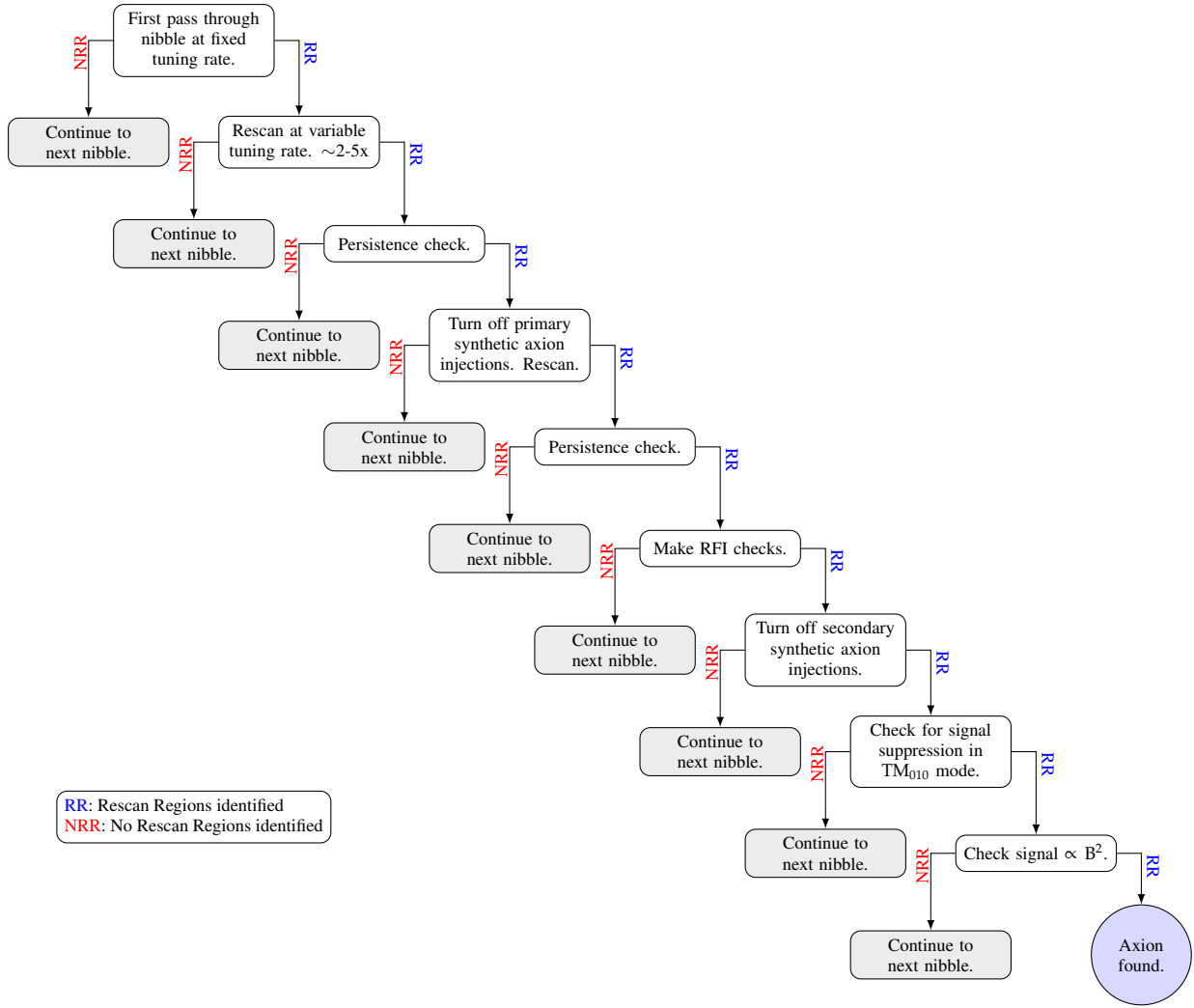

The haloscope technique is established as an effective means to search for axions, as evidenced by the fact that it is currently one of only a few types of experiment that have reached DFSZ sensitivity. Nevertheless, a well-known shortcoming of the haloscope technique is its inability to search over a wide range of axion masses \emph{quickly}. Therefore, a critical figure of merit for the axion haloscope is the \emph{scan rate}, which can be written as  
\begin{widetext}
\begin{multline}
    \frac{df}{dt}~{\approx}~157~\frac{\mathrm{MHz}}{\mathrm{yr}}\left(\frac{g_{\gamma}}{0.36}\right)^4\left(\frac{f}{740~ \mathrm{MHz}}\right)^2\left(\frac{\rho}{0.45\mathrm{~\mathrm{GeV/cm^3}}}\right)^2\left(\frac{3.5}{\text{SNR}}\right)^2\left(\frac{B}{7.6~\mathrm{T}}\right)^4\left(\frac{V}{136~\mathrm{\ell}}\right)^2\left(\frac{Q_{\text{L}}}{30{,}000}\right) \\ \left(\frac{C}{0.4}\right)^2\left(\frac{0.2~\mathrm{K}}{T_{\text{sys}}}\right)^2,
\end{multline}
\end{widetext}
\noindent where $T_{\text{sys}}$ is the system noise temperature~\cite{Sikivie1985,PhysRevD.36.974}. This equation represents the instantaneous scan rate; in other words, it does not account for ancillary measurements and amplifier biasing procedures. Data-taking operations involved tuning, with the scan rate set according to the parameters above. One advantage of a haloscope experiment, however, is that it possesses a robust means of confirming the existence of a dark matter candidate. The data-taking strategy for the run took the form of a decision tree such that advancement to each new step signified a higher probability of axion detection. The strategy is illustrated in Fig.~\ref{fig:decisiontree}.

The first step was tuning the cavity at a fixed rate over a pre-defined frequency range, called a \emph{nibble}, which was typically about 10-MHz wide, but varied depending on run conditions. Ideally, the first pass through a nibble would occur at a rate that was commensurate with achieving DFSZ sensitivity, although, due to fluctuating noise levels, that was not always the case.  The center frequency of spectra acquired under ideal operating conditions were typically spaced 2-kHz apart. The scan rate varied depending on the achievable operating conditions, including quality factor and system noise temperature.

Data-taking under these circumstances advanced as follows. Each 100-s digitization was accompanied by a series of measurements and procedures needed to characterize and optimize the receiver chain (Table~\ref{tab:cadence}). Every pass through this sequence was referred to as a single data-taking cycle and lasted approximately 2 minutes without JPA optimization.
\begin{table}
\setlength{\tabcolsep}{1pt}
\renewcommand{\cellalign}{lc}
\renewcommand{\arraystretch}{2.}
\begin{savenotes}
\setlength{\tabcolsep}{9pt}
\centering
\begin{tabular}{|l|l|l|l|}
\bottomrule[0.1ex]
\hline
Process & Frequency & \makecell{{Fraction of Time}\\{ per Iteration}} \\
\hline
\makecell{{Transmission}\\{Measurement}} &  \makecell{{Every}\\{Iteration}} & \makecell{~$<$ 1\%} \\
\hline
\makecell{{Reflection}\\{Measurement}} & \makecell{{Every}\\{Iteration}} & \makecell{~$<$ 1\%} \\
\hline
\makecell{{JPA}\\{Rebias}} &  \makecell{{Every 5-7}\\{Iterations}} & \makecell{~~25\%} \\
\hline
\makecell{{Check for}\\{SAG Injection}} &  \makecell{{Every}\\{Iteration}} & \makecell{~$<$ 1\%} \\
\hline
\makecell{Digitize} &  \makecell{{Every}\\{Iteration}} & \makecell{~~98\%\footnotemark} \\
\hline
\makecell{Move Rods} &  \makecell{{Every}\\{Iteration}} & \makecell{~$<$ 1\%} \\
\hline
\bottomrule[0.1ex]
\end{tabular}
\end{savenotes}
\footnotetext[1]{73\% for cycles when the JPA rebias procedure runs.}
\caption{Data-taking cadence. Ancillary procedures were used to characterize and optimize the RF system in real time. Axion search data were acquired only during the digitization process. SAG stands for synthetic axion generator, which was programmed to inject synthetic axions at specific frequencies.}
\label{tab:cadence}
\end{table}
An additional step of recoupling the antenna was also performed on occasion. This adjustment required user intervention and was done manually. Under ideal operating conditions, this cadence continued for the duration of a data `nibble', after which a rescan procedure was implemented. Rescans acquired more data in regions where axion candidates were flagged. The precise definition of what constitutes a candidate is described in Section~\ref{sec:rescan_procedure}. The rescan procedure used the same run cadence, but with significantly increased tuning rate, slowing down only at axion candidate frequencies. After rescan, all the data were examined to see if the candidate was persistent, followed by other tests to evaluate the axionic nature of the signal. The analysis was run continually throughout data-taking so that the scan rate could be adjusted in real time, to reflect changes in the experiment's sensitivity to axions. A detailed discussion of rescan procedure and data-taking decision tree can be found in Section~\ref{sec:rescan_procedure}.

\section{Analysis Inputs}
\label{section:noise_measurement}

\subsection{System Noise Characterization}

Central to any haloscope search is the ability to achieve a large SNR for axions. Given that ADMX operates in the high-temperature limit, where $hf\textless\textless k_{\text{B}}T$, the system noise temperature, $T_{\text{sys}}$, can be written as
\begin{equation}
T_{\text{sys}}=T_{\text{cav}}+T_{\text{amp}},
\label{eqn:tsys}
\end{equation}
\noindent where $T_{\text{amp}}$ is the noise temperature of the amplifiers and $T_{\text{cav}}$ is the physical temperature of the cavity. The amplifier noise can be written as
\begin{align}
\begin{split}
T_{\text{amp}}&=T_{\text{quantum}}+T_{\mathrm{HFET}}/G_{\text{quantum}}\\
&+T_{\text{post}}/(G_{\text{quantum}}G_{\text{HFET}}),
\end{split}
\end{align}
where $T_{\text{quantum}}$ is the noise temperature of the JPA, $T_{\mathrm{HFET}}$ is the noise temperature of the HFET, and $T_{\text{post}}$ is the noise temperature of the post-amplifier. The gain of the first stage amplifier (the JPA) is given by $G_{\text{quantum}}$, and the gain of the HFET is given by $G_{\mathrm{HFET}}$.

This means that the noise power, $P_{n}$  can be written as
\begin{equation}
    P_{n}=k_{\text{B}}T_{\text{sys}}b,
\end{equation}
\noindent where $P_{n}$ is the noise power, $k_{\text{B}}$ is the Boltzmann constant, and $b$ is the bandwidth over which the noise power is measured.  
The Dicke radiometer equation~\cite{doi:10.1063/1.1770483} in the high temperature limit provides the signal-to-noise ratio as 
\begin{equation}
    \text{SNR}=\frac{P_{\text{axion}}}{k_{\text{B}}T_{\text{sys}}}\sqrt{\frac{t}{b}},
\end{equation}
\noindent where $P_{\text{axion}}$ is the signal power of the axion.

Critical to quantifying the system noise temperature were measurements of the receiver temperature, which were acquired periodically throughout the course of the run. During Run 1B, four noise temperature measurements were made: one in February, one in July, one in September, and one in October of 2018. 
\begin{figure*}
    \centering
    \includegraphics[width=0.8\textwidth]{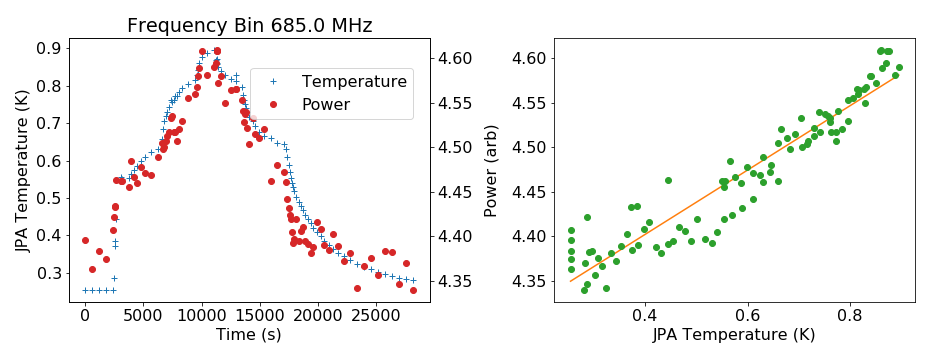}
    \caption{Heating the quantum amplifier package. The plot on the left shows the increase in quantum amplifier package temperature with time and power detected by the digitizer as a function of time during a $Y$-factor measurement of type 2. The plot on the right shows the power per unit bandwidth, measured off-resonance, as a function of temperature. The resulting fit, using Eq.~\ref{eqn:power_fit_equation_simple}, is shown in orange.}
    \label{fig:squidadel_heating}
\end{figure*}

In Run 1B, the receiver temperature had to be measured by halting ordinary data-taking operations and performing a $Y$-factor measurement~\cite{wilson2011techniques}. Although the goal of $Y$-factor measurement was to quantify the receiver noise temperature, there were two other unknowns that must be handled in this process: the attenuation between the cavity and the HFET, and the receiver gain. This information can be extracted via two different $Y$-factor techniques.

\subsection{$Y$-factor Method 1}

The first noise temperature measurement involved using the `hot load', labeled in Fig.~\ref{fig:run1b_receiver_chain}. Physically, the hot load consisted of an attenuator thermally sunk to a resistive heater. The hot load was connected to the switch via a superconducting NbTi coax line to minimize thermal conduction and attenuation. An excessive heat leak to the 4-K temperature stage limited the range over which the hot load temperature could be varied during this run.

An ideal noise temperature measurement would be performed with the JPA pump enabled, allowing the characterization of the noise along the entire receiver chain. However, the JPA does not maintain stable gain performance under changing temperatures and can saturate with small amounts of input thermal noise. Therefore, noise temperature measurements were performed with the pump disabled. The JPA, when turned off, was a passive mirror that allowed signals to propagate down the output line with minimal attenuation. 

Once the JPA was powered off, the RF switch was actuated so that the output RF line was connected to the hot load instead of the cavity. A heater and thermometer were attached to the hot load, enabling its temperature to be adjusted and measured. 

As the hot load was heated, a wide bandwidth power measurement was acquired by appending separate scans with roughly 5-MHz spacing. Under these conditions, the expected output power per unit bandwidth can be written as
\begin{equation}
P=G_{\text{HFET}}k_{\text{B}}\left[T_{\mathrm{JPA}}(1-\epsilon)+T_{\mathrm{load}}\epsilon+T_{\mathrm{HFET}}\right],
\label{eqn:power_fit_equation_load}
\end{equation} 
\noindent where $G_{\text{HFET}}$ is the HFET gain, $T_{\mathrm{JPA}}$ is the physical temperature of the quantum amplifier package and $T_{\mathrm{load}}$ is the physical temperature of the hot load. $T_{\mathrm{HFET}}$ is the noise attributed to the HFET and all downstream electronics, henceforth referred to as the \emph{receiver temperature}. The emissivity of the quantum amplifier package is given by $\epsilon$, which can be written as a function of the attenuation in the quantum amplifier package, $\alpha$: 
\begin{equation}
    \epsilon=10^{-\alpha/10}.
\end{equation}
Loss from the hot load to the JPA was quantified in two ways. First, it was quantified \emph{ex-situ} by measuring the losses in the two circulators. Next, it was quantified \emph{in-situ} using two different methods: first, by inferring it from a multi-component fit of a $Y$-factor measurement, and second by  using the JPA signal to noise ratio improvement (SNRI), and assuming that the JPA noise performance is independent of frequency. This was a reasonable assumption because variations in the noise performance are subdominant to other effects, such as variations in the circulator loss. The determination of the SNRI is discussed in the following section. A linear interpolation was used to increase the expected loss in the quantum amplifier package.

Equation~\ref{eqn:power_fit_equation_load} was then used on the $Y$-factor data to perform a two-component fit, where $G_{\text{HFET}}$ and $T_{\mathrm{HFET}}$ are fit parameters, whereas $T_{\mathrm{JPA}}$, $T_{\mathrm{load}}$, and $\epsilon$ were independently measured quantities. A hot load measurement of this type was performed twice throughout the course of Run 1B, on February 13 and October 9, 2018. 

\subsection{$Y$-Factor Method 2}

The other means of acquiring a receiver noise temperature measurement  was to apply a low enough voltage across the RF switch such that it would heat without flipping, thus, warming the quantum amplifier package. Noise temperature measurements of this type were performed on July 18 and September 12, 2018.

The power per unit bandwidth measured off-resonance by the digitizer in these configurations can be modeled with
\begin{equation}
P=G_{\text{HFET}}k_{\text{B}}\left[T_{\text{JPA}}(1-\epsilon)+T_{\text{cav}}\epsilon+T_{\text{HFET}}\right].
\label{eqn:power_fit_equation}
\end{equation}
\noindent Under these conditions, $T_{\mathrm{JPA}}$ is approximately equal to $T_{\mathrm{cav}}$, due to the location of the final stage attenuator, so that Eq.~\ref{eqn:power_fit_equation} simplifies to
\begin{equation}
    P=G_{\text{HFET}}k_{\text{B}}\left[T_{\text{JPA}}+T_{\text{HFET}}\right].
    \label{eqn:power_fit_equation_simple}
\end{equation}
This enables a separate confirmation of $T_{\mathrm{HFET}}$ that is independent of $\epsilon$. In this case, the fit parameters were the gain and $T_{\mathrm{HFET}}$, whereas $T_{\mathrm{JPA}}$ and $T_{\mathrm{cav}}$ were measured quantities. 
An example of such a measurement can be seen in Fig.~\ref{fig:squidadel_heating}. The left-hand side of Fig.~\ref{fig:squidadel_heating} shows the JPA temperature and power detected by the digitizer as a function of time. Over the course of the first 3 hours, a small voltage was incrementally increased to heat the hot load. The right-hand side shows the digitized power as a function of the JPA temperature, with the fit, using Eq.~\ref{eqn:power_fit_equation_simple}, shown in orange. There was no indication of any significant changes in the HFET over time, so the assumption was made that the HFET was stable throughout the course of the run.

\subsection{Combined Noise Temperature}
Both type 1 and type 2 $Y$-factor measurements were used to characterize the receiver noise temperature throughout the course of the run. The final analysis, however, relied on a combined receiver noise temperature measurement to set a limit. For Run 1B, it was realized that the receiver temperatures without the JPA taken throughout the course of the run did not vary significantly over the frequency range 680-760 MHz. This motivated the decision to generate a single noise temperature value that combined the results from our four measurements. The fit was achieved by calculating the expected residuals and the gain for each noise temperature measurement and performing a least squares fits on the combined result.  

\begin{figure}
    \centering
    \includegraphics[width=0.5\textwidth]{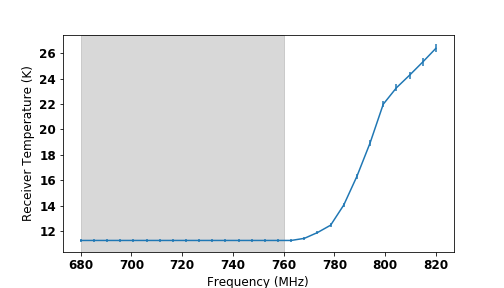}
    \caption{Combined receiver temperature over the frequency range for Run 1B. A noise temperature of 11.3 K\,$\pm$\,0.11 K was used from 680-760 MHz (highlighted in gray). The rise in the equivalent receiver temperature at the upper end of the frequency range is attributable to this range being the end of the circulator band.}
    \label{fig:receiver_temp}
\end{figure}

A plot of the combined receiver noise across the frequency range for Run 1B can be seen in Fig.~\ref{fig:receiver_temp}. The average value for the noise in the frequency range from 680 to 760 MHz was 11.3 K\,$\pm$\,0.11 K, where the error comes from the square root of the covariance from the fit. The receiver noise was higher at the upper end of the frequency range because of larger losses in the circulators near the end of the circulator band.

\begin{figure*}
    \centering
    \includegraphics[width=0.7\textwidth]{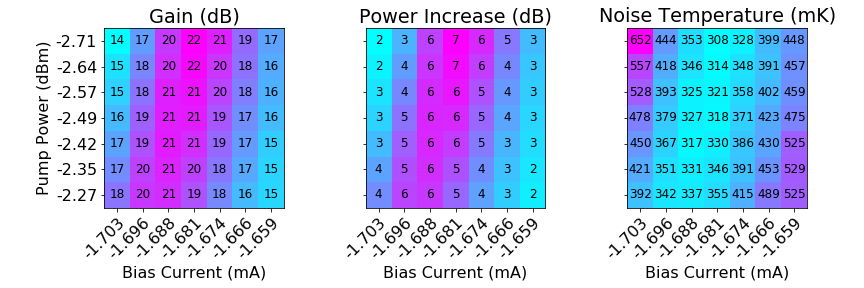}
    \caption{Sample SNRI calculated for several different bias and pump parameters during a single rebias procedure. \emph{Left:} Gain difference as measured by the network analyzer. \emph{Middle:} Increase in power as measured by the digitizer. \emph{Right:} The resultant noise temperature.}
    \label{fig:nice_snri_plot}
\end{figure*}

\subsection{SNRI Measurement}

The signal-to-noise ratio improvement (SNRI), commonly used to characterize quantum amplifiers, is defined as
\begin{equation}
\text{SNRI}=\frac{G_{\text{on}}}{G_{\text{off}}}\frac{P_{\text{off}}}{P_{\text{on}}},
\end{equation}
\noindent where $G_{\text{on}}$ is the gain with the JPA on, $G_{\text{off}}$ is the gain with the JPA off, $P_{\text{on}}$ is the measured noise power with the JPA on, and $P_{\text{off}}$ is the measured noise power with the JPA off.
The SNRI was monitored approximately every 10 min throughout the course of the run by measuring the gain and power coming from the receiver with the JPA on versus with the JPA off. This measurement occurred about once every 5-7 iterations through the full data-taking cycle. The SNRI typically did not vary more than 1 dB over this time frame. The SNRI changed throughout the course of the run because the JPA gain was not stable under changing temperatures; temperature variations on the order of 300-400 mK proved too large to guarantee gain stability. The HFET amplifier and upstream electronics were stable throughout the course of the run, so any SNRI changes could be attributed to the JPA. To mitigate any instability of the JPA, the SNRI was continuously optimized by searching over a range of pump powers and currents. A chart showing how the gain, power increase, and noise temperature vary with pump power and bias current at a given frequency is shown in Fig.~\ref{fig:nice_snri_plot}. Throughout data-taking, the JPA pump was offset by 375 kHz above the digitization region so as not to overwhelm the digitizer dynamic range.

\subsection{Total System Noise}
\label{sec:sys_noise}
\begin{figure}
    \centering
    \includegraphics[width=0.5\textwidth]{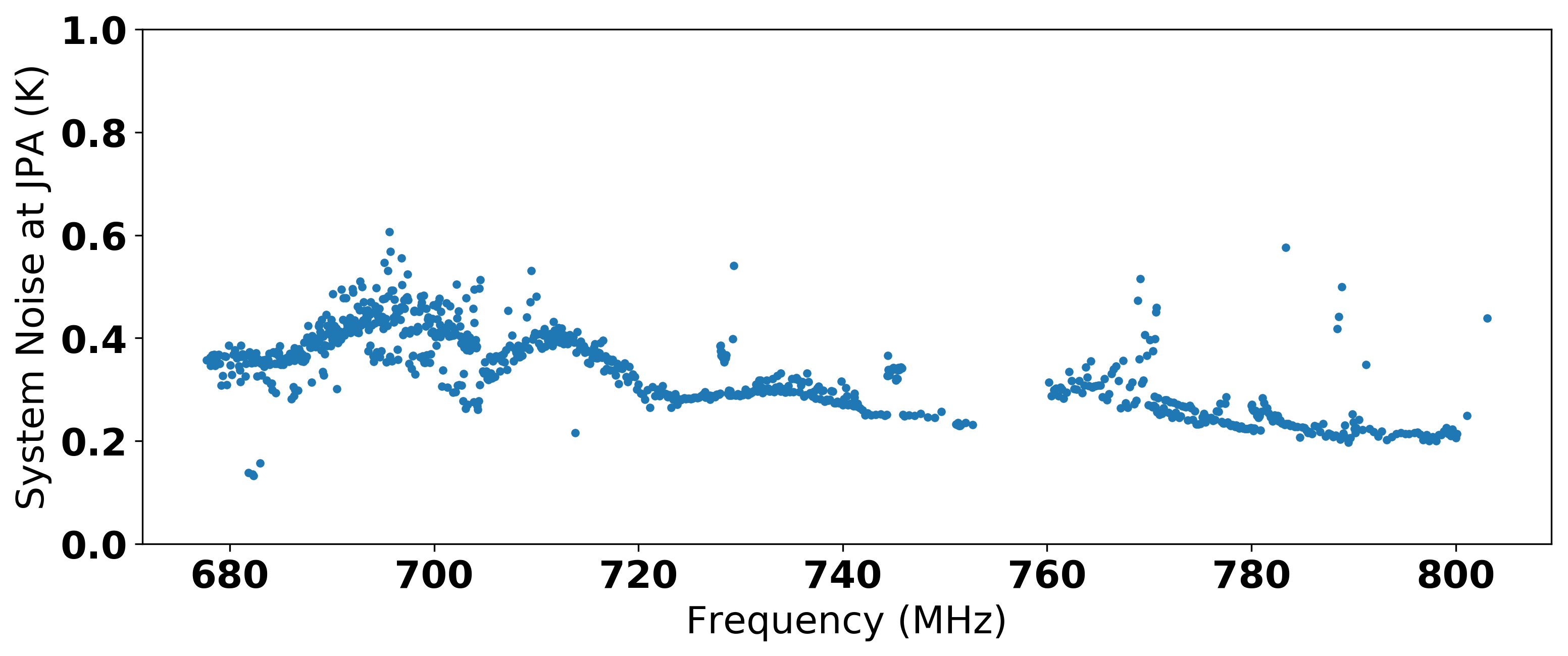}
    \caption{System noise as a function of frequency for the duration of the run.}
    \label{fig:tsys_vs_freq}
\end{figure}

The total system noise at the JPA input, given by Eq.~\ref{eqn:tsys}, can also be calculated from
\begin{equation}
T_{\mathrm{sys}}=T_{\mathrm{HFET}}/\mathrm{SNRI}.
\label{eqn:snr_improvement}
\end{equation}
A plot of the system noise at the JPA input over the full frequency range of Run 1B is shown in Fig.~\ref{fig:tsys_vs_freq}. To calculate the total system noise, one must account for the loss between the cavity and JPA as well. 

\subsection{Parameter Extraction}
Throughout the course of data-taking, ADMX tracked and monitored a number of system state data via various sensors and RF measurements. Data from temperature sensors were not tethered to the run cadence, whereas RF measurements typically occurred once per data-taking cycle (see Table~\ref{tab:cadence}).
The sensors were read out by numerous instruments, and the logging rate was a function of the capabilities and settings of a specific instrument. These instruments were queried every minute for their latest reading. To save memory, not every sensor reading was logged. Each sensor had a custom `deadband', or tolerance. If the preceding measurement was outside the `deadband', the sensor would be logged. If, after 10 minutes, no sensor readings existed outside the `deadband', the sensor reading would be logged regardless. 

Aside from the SNRI rebiasing procedure, RF measurements occurred once every data-taking cycle. The following parameters were extracted from these measurements to be used in the analysis:

\begin{enumerate}
    \item Quality factor as measured by transmission scans.
    \item Resonant frequency as measured via transmission scans.
    \item Coupling coefficient (which can be thought of as the ratio of the impedance of the cavity and the impedance of the 50-ohm transmission line connected to the cavity), as measured via reflection scans.
\end{enumerate}
\noindent The cavity coupling coefficient, $\beta$, was given by 
\begin{equation}
	\Gamma=\frac{\beta-1-(2iQ_0\delta\omega/\omega_0)}{\beta+1+(2iQ_0\delta\omega/\omega_0)},
\end{equation}
where $\Gamma$ is the reflection coefficient of the cavity, $Q_0$ is the unloaded quality factor, $\omega$ is the frequency, and $\omega_0$ is the resonant frequency. Using this equation, a fit to the coupling constant, $\beta$, was performed on the complex and imaginary data obtained from a reflection measurement~\cite{brubaker2018results,pozar2009microwave}.

Since the quality factor, resonant frequency and coupling were expected to change very slowly with frequency, more accurate measurements could be obtained by smoothing. The coupling coefficient was smoothed over a period of 30 min, whereas the quality factor was smoothed over a period of 15 min. Neither the quality factor nor the coupling parameters varied significantly over these time scales.

The form factors were simulated and read in from a separate file. The simulation used the Computer Simulation Technology (CST) software~\cite{CST}. The output of the simulation was the form factor at a few select frequencies; to acquire a form factor at every point in frequency space, the simulated data were interpolated.
\begin{figure}
    \centering
    \includegraphics[width=0.5\textwidth]{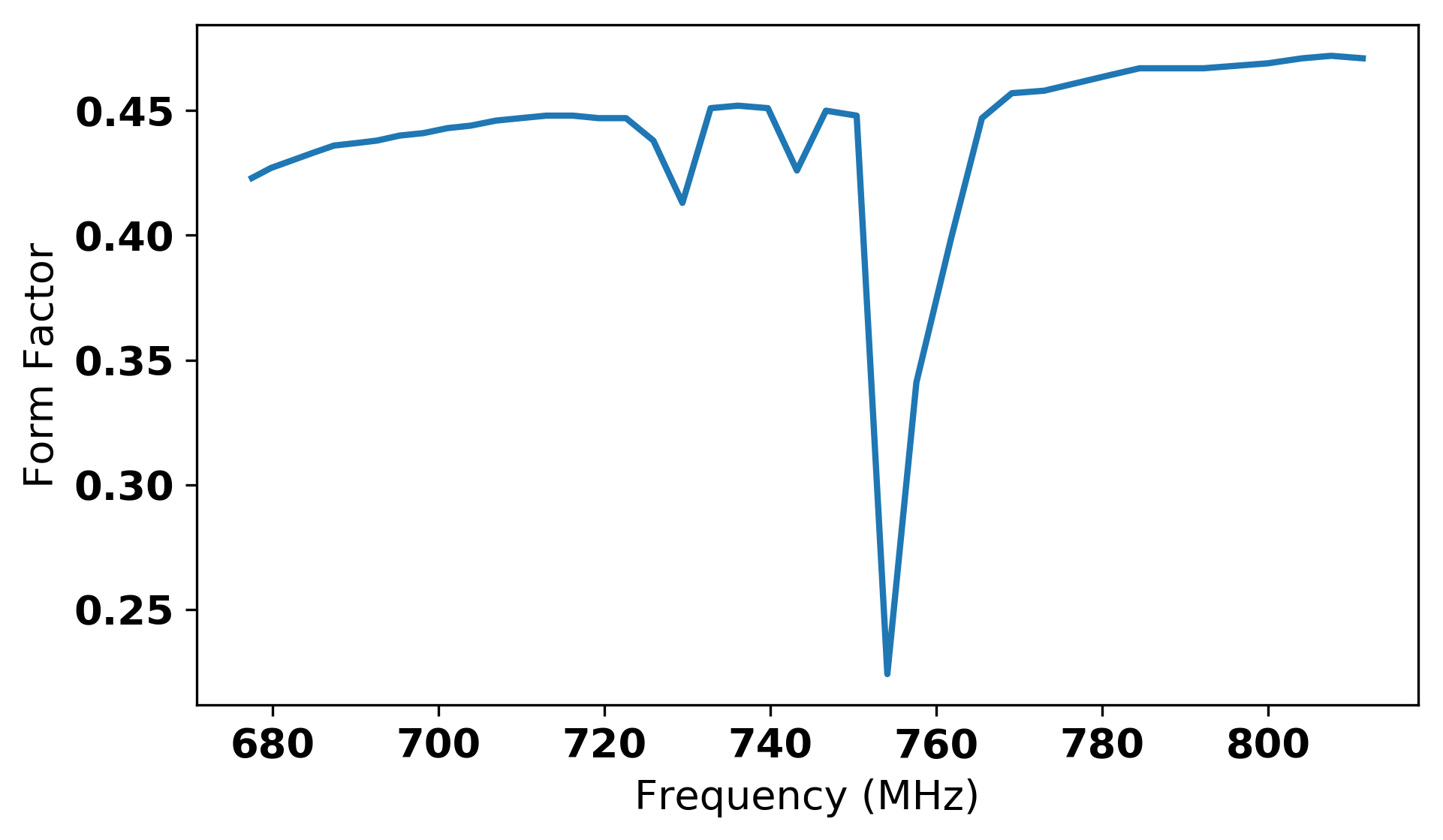}
    \caption{Form factor as a function of frequency. The dip near 750 MHz is at the location of mode crossings.}
    \label{fig:formfactors}
\end{figure}

The system noise across the full frequency range for Run 1B, as described in Section~\ref{section:noise_measurement}, was also provided as an input to the analysis. The system noise was composed of the receiver temperature divided by the SNRI and the loss between the cavity and the HFET amplifier. While the SNRI was interpolated in time, the receiver temperature was interpolated at each point in frequency space.

\subsection{Systematics}

The systematic uncertainty was quantified for the following parameters that were used in the analysis. A summary of all systematics can be seen in Table~\ref{tab:uncertainties}.

First, the uncertainty in the quality factor was quantified by repeatedly measuring the quality factor in a narrow range of frequencies, 739-741 MHz, where the quality factor was not expected to change much as a function of frequency, according to models. The fractional uncertainty in the quality factor in this range was determined to be $\pm$\,1.1\%. The fractional uncertainty in the coupling was also computed over the same frequency range, and determined to be $\pm$\,0.5\%. The fractional uncertainty from the $Y$-factor measurements is cited as `RF model fit', and accounts for uncertainty in the receiver temperature as well as the uncertainty in the attenuation. Uncertainty on our temperature sensors came from the values stated on their datasheets. This factored into the uncertainty of the receiver noise temperature, and therefore the system noise.

The uncertainty in the SNRI measurement was evaluated using the following method. It was observed that the measured SNRI varied as a function of the JPA gain, with the worst uncertainty occurring at high JPA gain. The largest observed uncertainty was $\pm$\,0.18 dB, corresponding to a linear uncertainty of $\pm$\,0.042 in the power measured in each bin of the grand spectrum.


\begin{table}
\centering
\renewcommand{\arraystretch}{1.3} 
\begin{tabular}{@{}lc@{}}
    \toprule[0.1ex]
    \hline
     Source & Fractional Uncertainty  \\
     \toprule[0.1ex]
     \hline
     $B^2VC_{\text{010}}$ & 0.05 \\
     $Q$ & 0.011 \\
     Coupling & 0.0055 \\
     RF model fit & 0.029 \\
     Temperature Sensors & 0.05 \\
     SNRI measurement & 0.042 \\
     \bottomrule[0.1ex]
     \hline
     Total on power & 0.088 \\
     \hline
     \bottomrule[0.1ex]
     \hline
\end{tabular}
\caption{Dominant sources of systematic uncertainty. The uncertainties were added in quadrature to attain the uncertainty on the total axion power from the cavity, shown in the bottom row. For the first entry, $B$ is the magnetic field, $V$ is the volume, and $C_{\text{010}}$ is the form factor. The last row shows the total uncertainty on the axion power from the cavity.}
\label{tab:uncertainties}
\end{table}

The total systematic uncertainty of $\pm$\,0.088, shown in Table~\ref{tab:uncertainties}, was computed simply by adding all listed uncertainties in quadrature.

\section{Axion Search Data-Processing}
\label{sec:analysis_procedure}
\subsection{Baseline Removal}

The first step in processing the raw spectra was to remove the fixed baseline imposed on the spectra from the warm electronics. A nonflat power spectrum had three possible underlying causes:
\begin{enumerate}
    \item Frequency dependent gain variations after mixing.
    \item Frequency dependent gain variations before mixing.
    \item Frequency dependent noise variations.
\end{enumerate}

\begin{figure}
    \centering
    \includegraphics[width=0.5\textwidth]{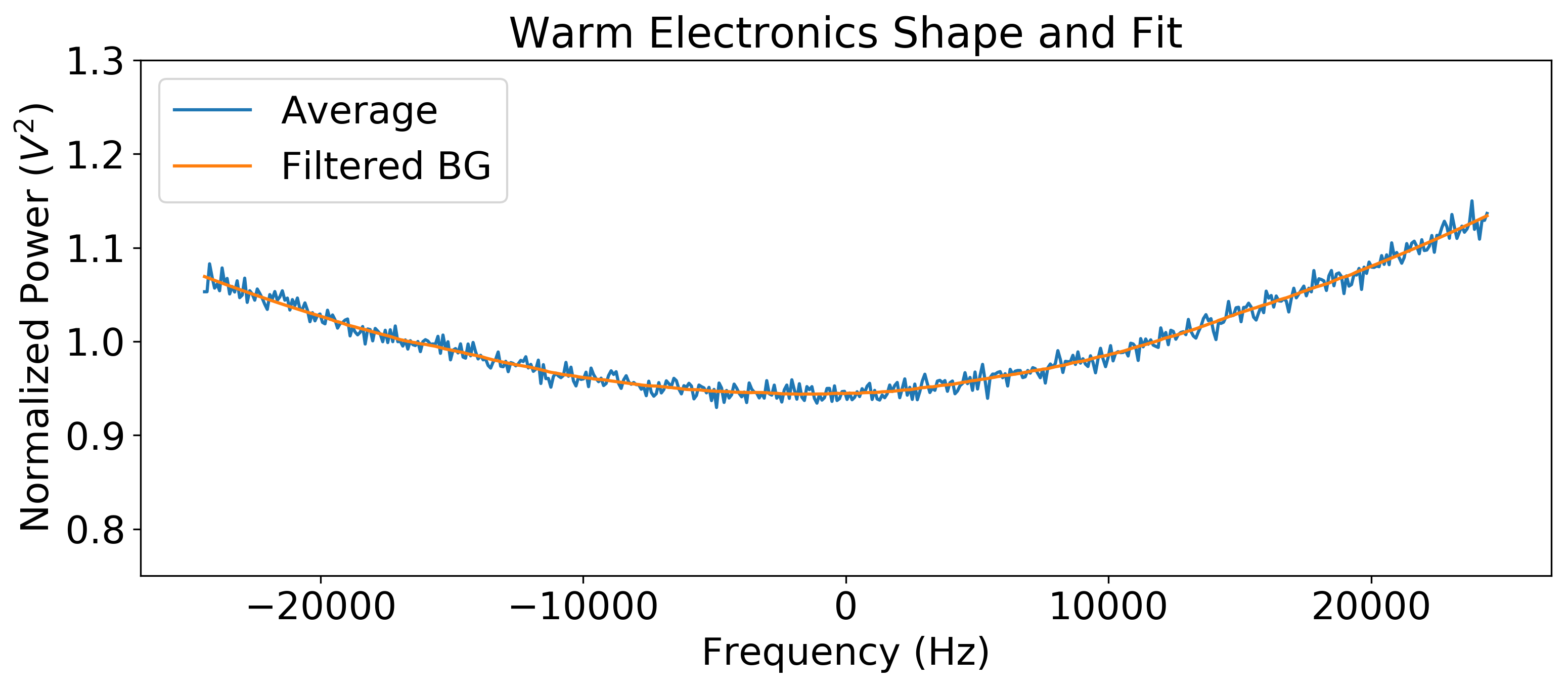}
    \caption{Filtered background shape (`Filtered BG' and orange) and the average baseline (`Average' and blue) from the warm electronics.}
    \label{fig:bg}
\end{figure}

The last of these was subdominant because most noise sources had approximately the same temperature. Gain variations before mixing, attributable to interactions of RF devices in the cold space, were evident, but small compared to gain variations after mixing. Gain variations after mixing were primarily determined by filters in the receiver chain. The characteristic shape of these gain variations, also known as the spectrum's \emph{baseline}, can be seen in Fig.~\ref{fig:bg}. The upwards trends to the far right and left were a result of digitizing in the final two-pole 150-kHz bandpass filter, between the two poles. The baseline was averaged and smoothed using a Savitzky-Golay software filter ~\cite{SavGol,Malagon:2014nba}.

The average baseline is shown in blue and the filtered background is shown in orange. The y-axis was normalized because the original scale is arbitrary and a combination of the gain and attenuation of the output line.

\subsection{Spectrum Processing}

\begin{figure}
\begin{minipage}[t]{0.5\textwidth}
\includegraphics[width=\textwidth]{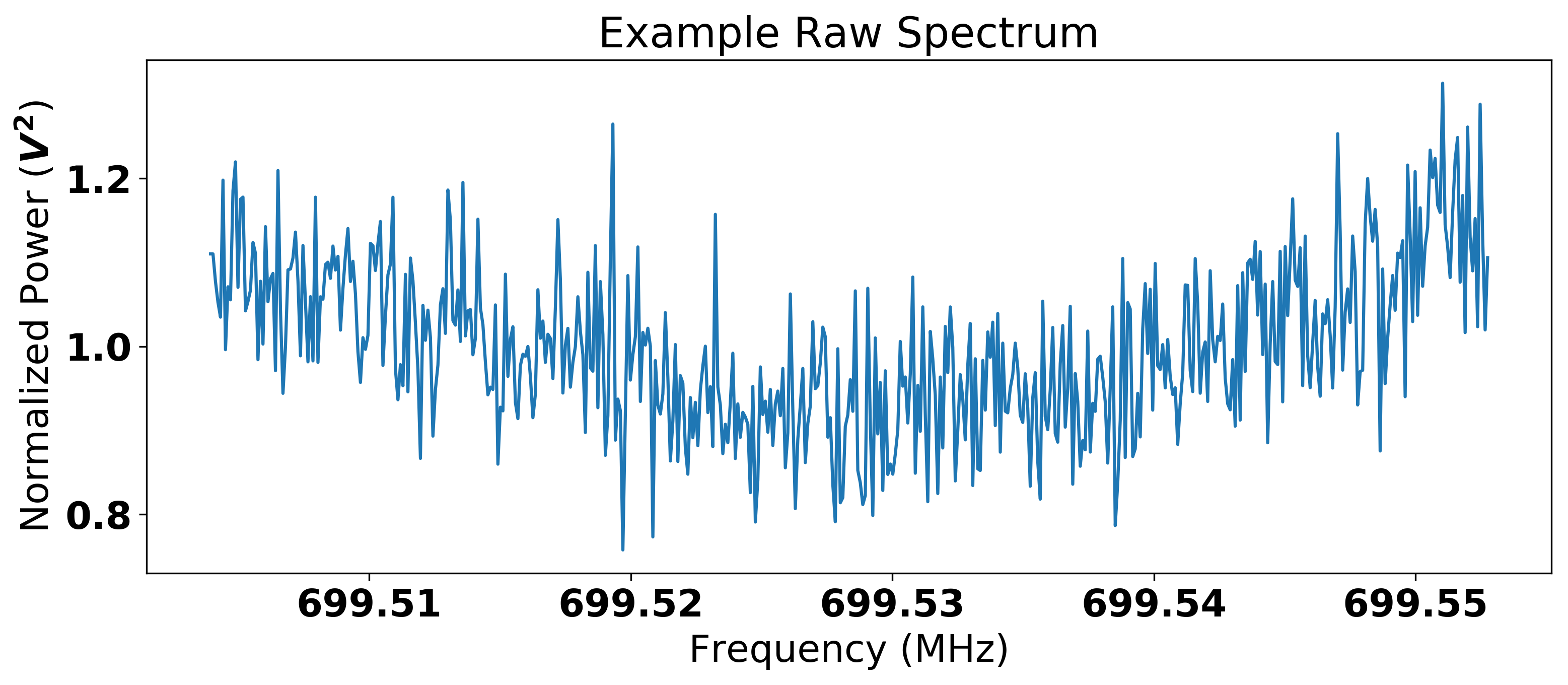}
\caption{Raw spectrum, or single digitizer scan. All the raw scans have a distinct shape imposed by the receiver chain.}
\label{fig:raw_spectrum}
\end{minipage}
\end{figure}

\begin{figure}
    \includegraphics[width=0.5\textwidth]{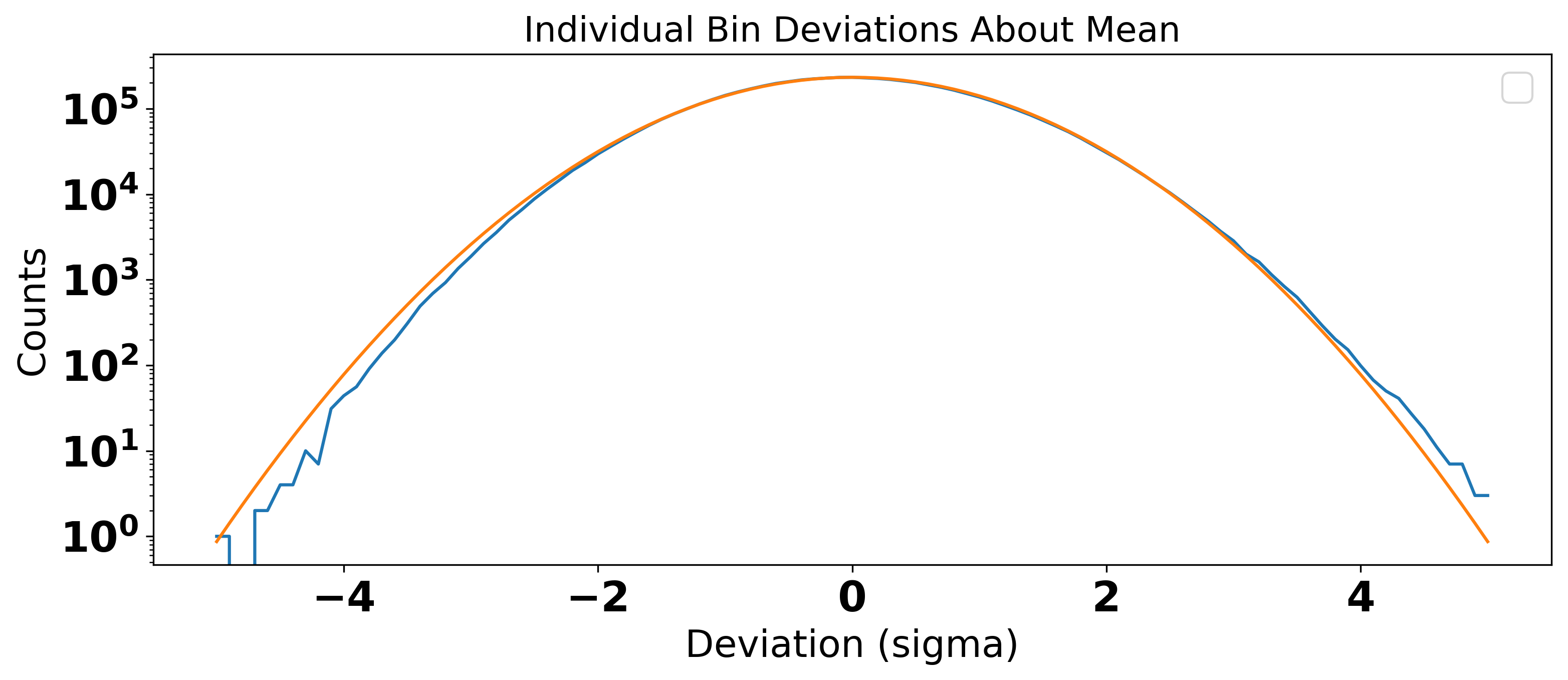}
    \caption{Histogram (blue) of individual bin deviations about the mean for the first nibble of Run 1B. The orange curve is a Gaussian fit to the data.}
    \label{fig:whitenoise}
\end{figure}

\begin{figure}
    \includegraphics[width=0.5\textwidth]{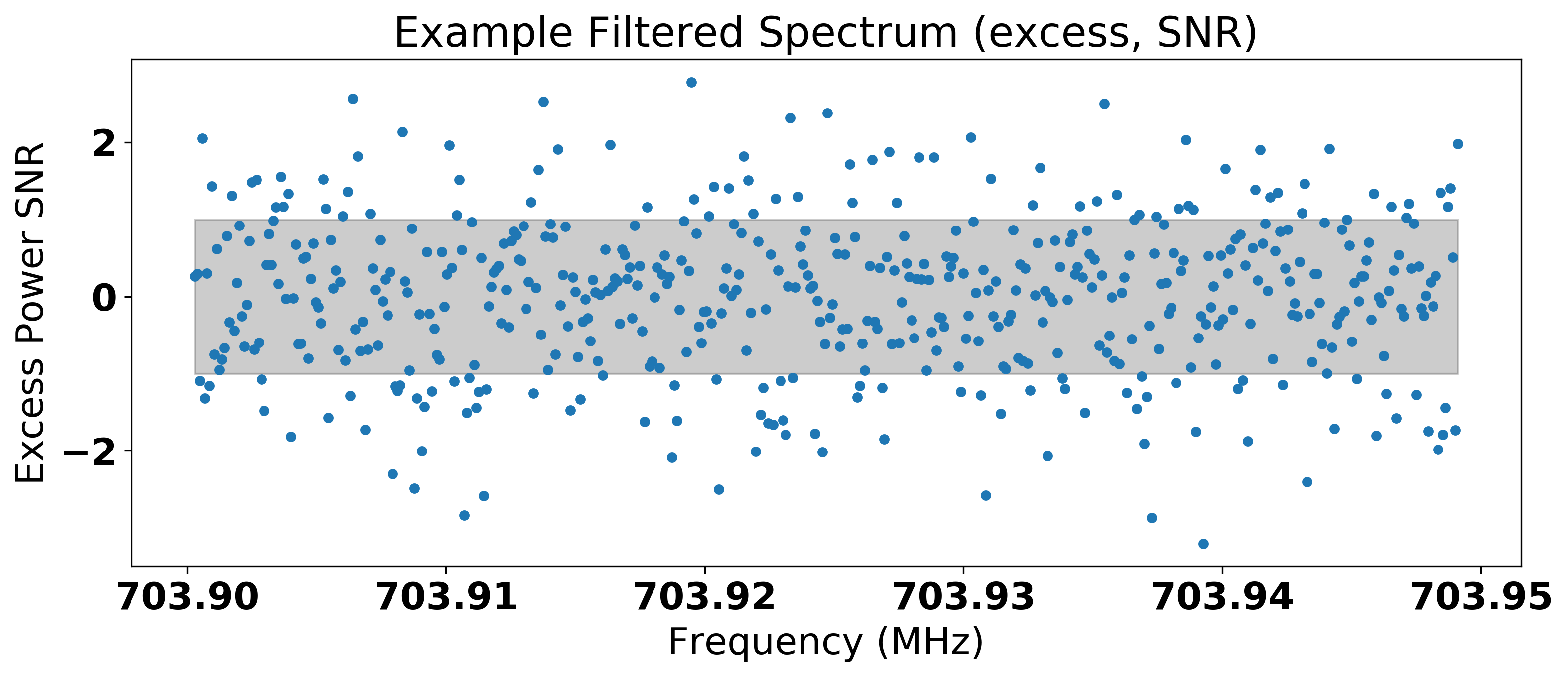}
    \caption{Filtered Spectrum.}
    \label{fig:filter_spectrum}
\end{figure}

An example spectrum after the baseline removal procedure is shown in Fig.~\ref{fig:raw_spectrum}. Each raw spectrum consisted of 512 bins, with bin widths of 95 Hz, for a total spectrum width of 48.8 kHz. A single spectrum is representative of axion search data acquired over an integration time of 100 s, a combination of $\mathrm{10^4}$ Fourier transforms of 10 ms of cavity output signal. In the following discussion, the smallest discretization of measured power is defined by $P^{j}_{i}$, where $j$ identifies an individual spectrum, and $i$ identifies an individual bin. Each raw spectrum was processed individually as follows. First, the raw power was divided by the baseline and convolved with a six-order Padé filter to remove the residual shape from the cryogenic receiver transfer function. The use of a Padé filter was motivated by deriving the shape of the power spectrum at the output of the last-stage cold amplifier~\cite{Daw:2018tru}. 
The power in each bin was then divided by the mean for the entire spectrum to create a normalized spectrum. In the absence of an axion signal, the power in each bin could then be represented as a random sample from a Gaussian distribution with a mean of $\mu=1$. Evidence that this was indeed the case can be seen in Fig.~\ref{fig:whitenoise}, where a Gaussian fit to the data is shown in orange. Subtracting 1 from each bin shifted the mean of the normalized spectrum to $\mu=0$, which gave a more intuitive meaning to the data, enabling us to search for power fluctuations above zero. An example of such a filtered spectrum is shown in Fig.~\ref{fig:filter_spectrum}. The gray band highlights the 1$\sigma$ error bar, which implies 68\% of the data falls within this region. 

\begin{figure}
    \includegraphics[width=0.5\textwidth]{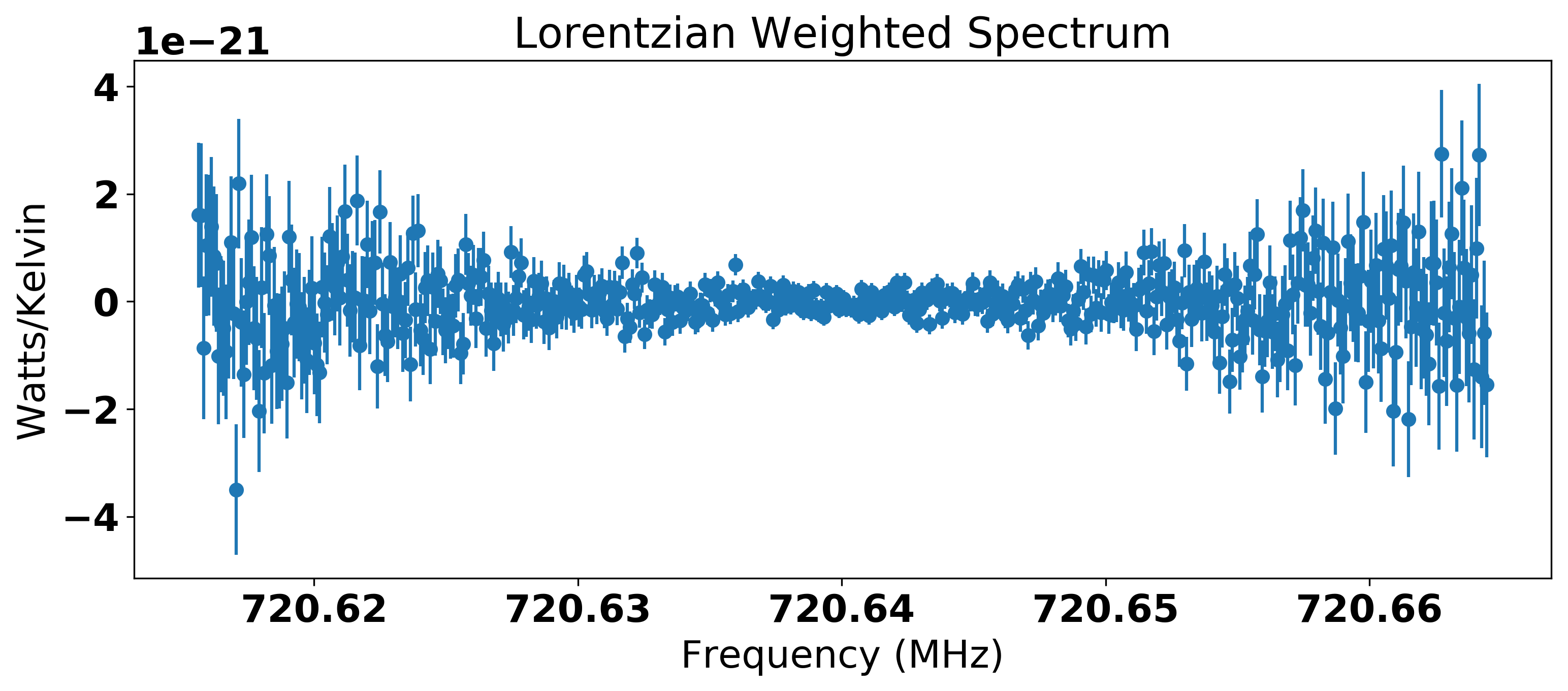}
    \caption{Lorentzian weighted spectrum divided by the noise power}
    \label{fig:lor_spectrum}
\end{figure}

Another feature of the raw data that must be considered is inherited from the microwave cavity itself: the Lorentzian shape. Power measured closer to cavity resonance is enhanced by the full $Q$ of the cavity, whereas power measured further from resonance is not. The enhancement follows the Lorentzian shape of the cavity, which varies depending on the coupling and frequency at the the time of the scan. The filtered spectra were therefore scaled by their respective Lorentzian shapes. The result of this step can be seen in Fig.~\ref{fig:lor_spectrum}, where the error bars are indicative of the distance from the cavity resonance peak. 

\subsection{Implementation of Analysis Cuts}

Five analysis cuts, shown in Table~\ref{tab:analysis_cuts}, were applied for quality control of the data. The original Run 1B data consisted of 197,680 raw spectra. After implementing the analysis cuts shown in Table~\ref{tab:analysis_cuts}, 185,188 raw spectra remained. Motivation for these cuts proceeded as follows. First, quality factors lower than 10,000 and greater than 120,000 were omitted from the data because they were likely unphysical and the result of a poor fit to a noisy transmission measurement. System noises below 0.1 K and above 2.0 K were excluded, as these were likely simply to be the result of incorrectly measuring the SNRI. Temperatures below 0.1 K were removed because they were lower than any physical temperature in the experiment and would violate the Standard Quantum Limit. Additionally, the six-order Padé fit to the background was required to have a ${\chi}^2$ per degree of freedom less than 2. This proved sufficient enough to reject poor fits while retaining potential axion signals. 

In addition to these parameter cuts, cuts were also made over various time stamps as a result of aberrant run conditions. Reasons included uncoupling of the antenna, digitizer failures, software malfunctions, excursions of the SNR that required better background fitting, scans containing pervasive and obvious RFI, unexpected mode crossings, a poorly biased JPA, and various engineering studies. These studies ranged from manual rebiasing of the JPA, to heating or cooling of the dilution refrigerator, to ramping the main magnet.

\begin{center}
\begin{table}
\centering
\renewcommand{\arraystretch}{1.3} 
\begin{tabular}{@{}lcr@{}}
\toprule[0.1ex]
\hline
\toprule[0.1ex]
Cut Parameter & Scans Removed & Constraint \\
\hline
Timestamp cuts & 7,189 & N/A \\
\hline
 Quality Factor & 316 & 10,000
 \textless\,$Q$\,\textless 120,000 \\ 
 \hline
 System Noise & 4,514 & 0.1 \textless\,$T_{\text{sys}}$\,\textless 2.0  \\  
 \hline
 Max Std. Dev. Increase & 224 & 2.0 \\   
 \hline
 Error in filter shape & 249 & N/A \\
 \hline
 \bottomrule[0.1ex]
 \hline
\end{tabular}
\caption{Table of analysis cuts made to spectra.}
\label{tab:analysis_cuts}
\end{table}
\end{center}

\subsection{Grand Spectrum Preparation}

The final step of the analysis was to merge all the power spectra into a single grand spectrum. This presents a challenging problem: how does one combine overlapping spectra into a single RF bin? The conditions under which each scan is acquired change, and so each must be weighted accordingly. 
The primary priority of such an endeavor is to control for these varying conditions throughout the run. As in previous analyses, the way this was accomplished was to scale the power excess in each bin of the normalized spectrum by the power that would be generated by a DFSZ axion signal (see Eq.~\ref{eqn:axion_pwr}) under the conditions present during that particular scan acquisition (inputting the measured $Q$, $f$ and $C(f)$ for that scan). 


Another condition that must be controlled between spectra is the system noise. All else considered, axion peaks of identical signal power but different noise temperatures lead to different peak heights. By scaling each bin in the normalized spectrum by the noise power, $k_{\text{B}}T_{\text{sys}}$, the effects of varying system noise are mitigated. 

Scaling by the axion signal power parameters and accounting for the differences in system noise requires computing 
\begin{equation}
    P^{j}_{i_\text{scaled}}=P^{j}_{i,\text{lor}}\left(\frac{1}{C_{010}}\right)\left(\frac{1~\mathrm{m^3}}{V}\right)\left(\frac{1}{Q}\right)\left(\frac{1~\mathrm{T^2}}{B^2}\right)
\end{equation}
\noindent on a bin-by-bin basis, where $P^{j}_{i,\text{lor}}$ is the filtered power from an individual frequency bin and spectrum, scaled by the Lorentzian shape of the cavity.  The effect of all this processing is to remove all possible discrepancies between scans, enabling apples-to-apples comparisons of the power between bins, resulting in $P^{j}_{i_\text{scaled}}$.

To further increase sensitivity to a potential axion signal, one final step is performed before combining the data into a grand spectrum: filtering in accordance with the axion lineshape. It is well known that an axion signal would have a characteristic lineshape reflective of the axion kinetic energy distribution~\cite{PhysRevD.42.3572}. The velocity of axions in the case of an isothermal, virialized halo would follow a Maxwell-Boltzmann distribution. This distribution derives from the assumption that dark matter obeys the standard halo model (SHM), which describes the Milky Way Halo as thermalized, with isotropic velocity distribution. The Maxwell-Boltzmann lineshape is 
\begin{equation}
    g(f)=\frac{2}{\sqrt{\pi}}\sqrt{f-f_{a}}\left(\frac{3}{f_{a}\frac{c^2}{\braket{v^2}}}\right)^{3/2}e^{\frac{-3(f-f_a)}{f_a}\frac{c^2}{\braket{v^2}}},
    \label{eqn:MaxBoltz}
\end{equation}
\noindent where $f$ is the measured frequency and $f_a$ is the axion rest mass frequency. The rms velocity of the dark matter halo is given by $\braket{v^2}=(270\;\mathrm{km/s})^2$~\cite{PhysRevD.42.3572}.
The measured power in each bin is then convolved with a Maxwell-Boltzmann filter which uses this distribution. Note that the effects from the orbital motion of Earth around the Sun and the rotational motion of the detector about the axis of the Earth have been averaged out in this equation. The medium-resolution analysis does not have the required spectral resolution to observe the Doppler effect of such motion, which would result in a frequency shift that can be attributed to daily and yearly modulation. A separate, `high-resolution' analysis is underway which would be capable of detecting this shift. Additionally, at this stage of the analysis, an alternative axion velocity distribution, known as an N-body lineshape, was be used as a filter. This filter emerged from developments in galaxy formation simulations for the Milky Way. The simulation describes galaxies using the N-body+smooth-particle-hydrodynamics (N-Body+SPH) method, in lieu of the assumption that the dark matter obeys the Standard Halo Model (SHM). The N-body signal shape keeps a Maxwellian-like form
\begin{equation}
    g({f})\approx\left(\frac{(f-f_{a})}{m_{a}\kappa}\right)^{\alpha}e^{-\left(\frac{(f-f_{a})}{m_{a}\kappa}\right)^{\beta}}.
\end{equation}
\noindent The best fit parameters were computed via simulation, and found to be $\alpha = 0.36~{\pm}~0.13$, $\beta = 1.39~{\pm}~0.28$, and $\kappa=(4.7~{\pm}~1.9){\times}10^{-7}$~\cite{0004-637X-845-2-121}. The medium resolution analysis results were also computed separately with this filter which produced different limits on the axion coupling relative to the assumption of a Maxwell-Boltzmann distribution.

Combining individual spectra into a grand spectrum involves the use of a well-established `optimal weighting procedure'~\cite{Brubaker:2017rna,PhysRevD.64.092003}. The optimal weighting procedure finds weights for the individual power excesses that result in the optimal SNR for the grand spectrum~\cite{Brubaker:2017rna}. In this procedure, the weights are chosen such that the maximum likelihood estimation of the true mean value, $\mu$, is the same for all contributing bins. 

More rigorously, the grand spectrum power excesses can be computed on a bin-by-bin basis using the following equation
\begin{equation}
   P_{w}=\frac{\sum\limits_{j=0}^{N} \frac{P^{j}_{scaled}}{{\sigma^{j}}^2}}{\sum\limits_{j=0}^{N}\frac{1}{{\sigma^{j}}^2}},
\end{equation}
\noindent where $N$ is the total number of spectra for a given frequency bin, and $P_{w}$ is the weighted power for an individual RF bin of the grand spectrum. The standard deviation for each bin in the grand spectrum is calculated via
\begin{equation}
\sigma_{w}=\sqrt{\frac{1}{\sum\limits_{j=0}^{N}\frac{1}{{\sigma_{j}}^2}}}.
\end{equation}
The grand spectrum is completely defined, bin-by-bin, by these two values: the measured excess power, $P_{w}$, and the standard deviation, $\sigma_{w}$.

Searches for excess signals above the noise in the grand spectrum that would correspond to an axion are further delineated in the following sections.

\section{Synthetic Axions}
\label{sec:synthetics}
There were two types of synthetic axions signals used in Run 1B: software and hardware synthetic injections. Synthetics were used to build confidence in our analysis. 

\subsection{Software Synthetics}

Software synthetics serve the purpose of better understanding the analysis---in particular, the detection efficiency. Software synthetics reflected the axion lineshape as described by the Maxwell-Boltzmann distribution, and their power levels could be adjusted relative to the KSVZ axion power. By injecting 1,773 evenly-spaced software signals at DFSZ power with a density of 20 per MHz into the real data and checking what fraction were flagged as candidates, the detection efficiency was calculated. This process was performed for all data that were collected at DFSZ sensitivity. Of these 1,773 injections, 1,684 were detected, corresponding to a $95\pm2\%$ detection efficiency of DFSZ signals.

\begin{figure}
    \centering
    \includegraphics[width=0.5\textwidth]{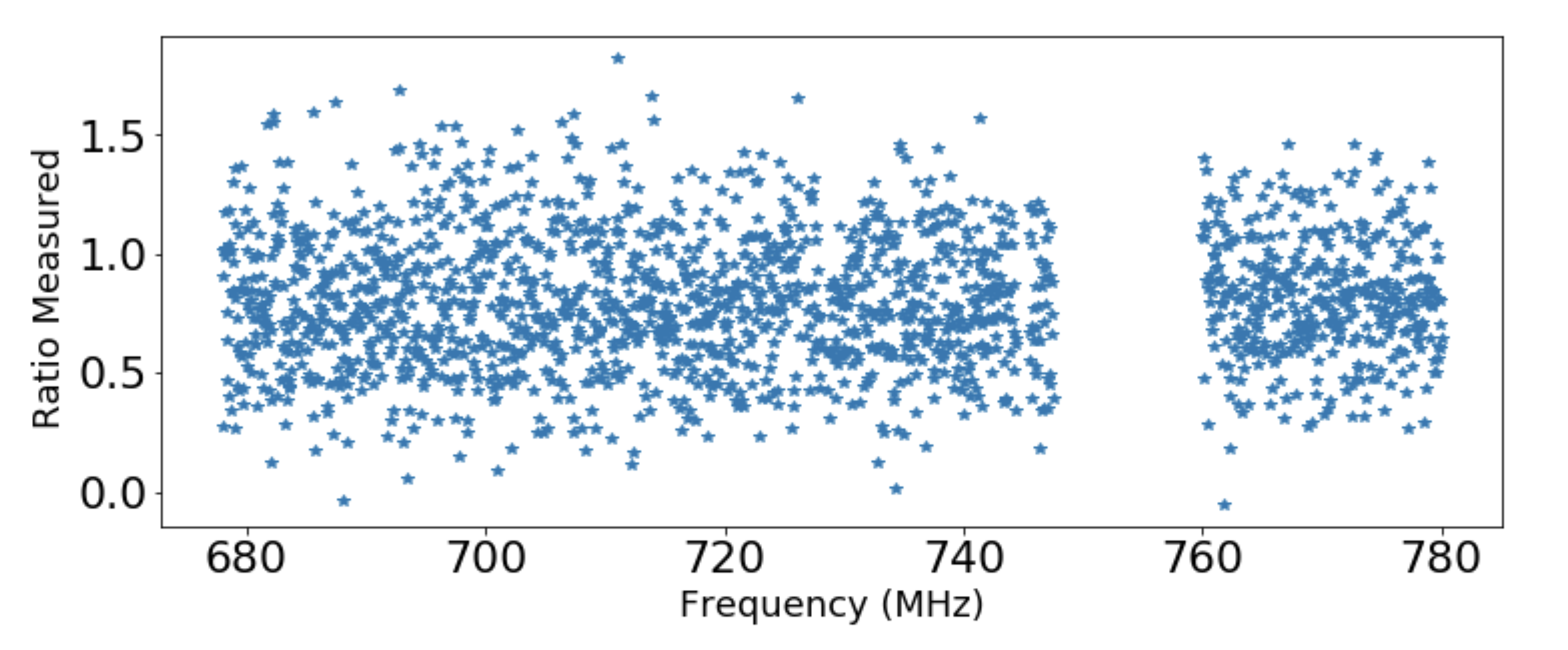}
    \caption{Ratio of measured power to injected software synthetic power over the full frequency range for Run 1B (the gap at 750-760 MHz was a large set of mode crossings).}
    \label{fig:fudgecalc}
\end{figure}

It was also discovered that the background fit can reduce the significance of axion signals. This effect arises from the fact that the background fit is designed to accurately describe wide features and ignore narrow peaks so as not to accidentally fit out a potential axion candidate. This effect was quantified by calculating the ratio of the power of the injected synthetic signal to the measured power. This ratio was computed across the relevant frequency range and can be seen in Fig.~\ref{fig:fudgecalc}. The average ratio was $0.818\,\pm\,0.008$. Power measurements from the grand spectrum were therefore corrected by dividing by this ratio to account for sensitivity loss from the background fit or other analysis steps.

\subsection{Hardware Synthetics}
\begin{figure}
    \centering
    \includegraphics[width=0.5\textwidth]{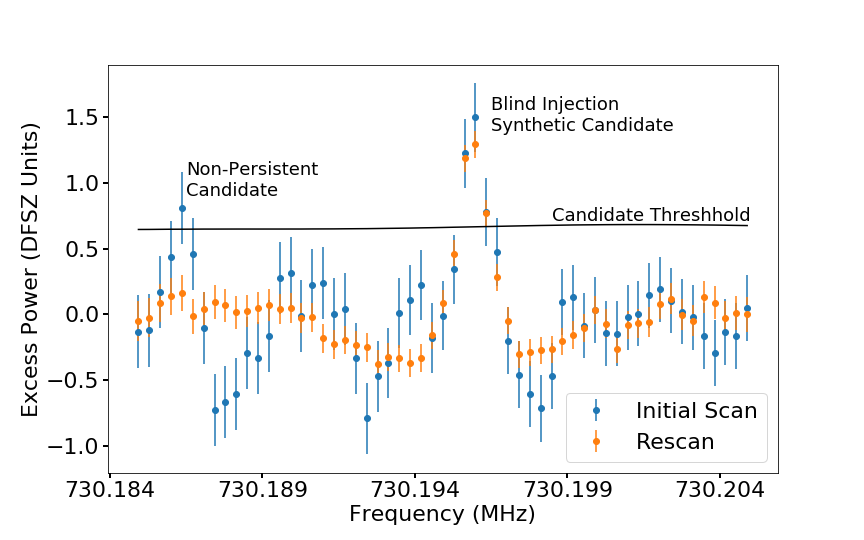}
    \caption{Hardware synthetic injection. Blue shows the results from the initial set of scans over this frequency interval, and orange shows the results after rescans (with the synthetic candidate still present).}
    \label{fig:hardware_synth}
\end{figure}

The hardware synthetic axions were a novel addition to ADMX for Run 1B and were used for better understanding of the receiver chain and sensitivity. The Synthetic Axion Generator (SAG) was located in a separate rack, away from ordinary data-acquisition. The SAG consisted of an arbitrary waveform generator (Agilent 33220A) that created a low frequency Maxwell-Boltzmann-like signal, about 500-Hz wide. This signal was mixed up to a specific RF frequency and injected into the cavity via the weak port as it was tuned through that frequency. The attenuation was calibrated by intentionally injecting synthetic axions of known attenuation and measuring their output power, so that signals could be sent in during the run as fractions of DFSZ signal power. Hardware synthetics were injected into the weak port of the receiver chain via a blind injection scheme throughout the course of the run. These synthetics were successfully detected, confirming our understanding of the receiver chain and analysis. An example of such a synthetic axion that was detected and flagged as a candidate via the analysis is shown in Fig.~\ref{fig:hardware_synth}.
\section{Mode Crossings}
\label{sec:mode_crossings}
\begin{figure*}
\centering
\includegraphics[width=0.8\textwidth]{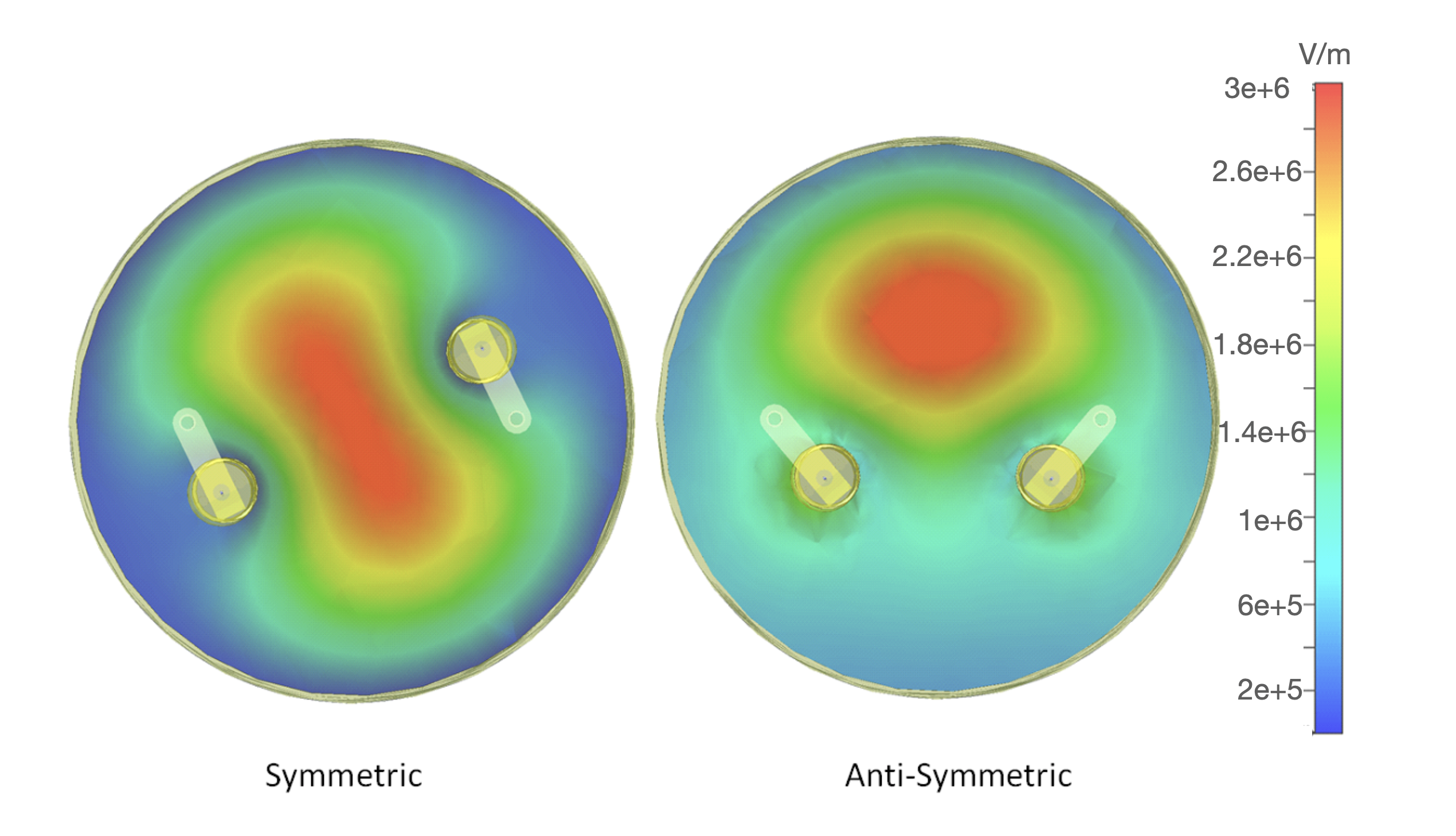}
\caption{Positioning of the rods in symmetric vs. anti-symmetric configurations. Normal data-taking operations used the symmetric mode (both rods moving counter-clockwise to brings rods to the center and increase frequency) whereas the anti-symmetric mode (left rod moving counter-clockwise with right rod moving clockwise to bring both rods to the center and increase frequency) was used to navigate mode crossings. The color scale shows the electric field strength (V/m) as modeled by CST Microwave Simulation~
\cite{CST}.}
\label{fig:rod_motion}
\end{figure*}

The original axion search for Run 1B proceeded with the tuning rods operating in what is known as the `symmetric configuration'. That is to say, the rods, starting at the same position opposite each other next to the walls, were rotated in the same direction, at the same rate. The first pass through the Run 1B frequency band included 8 mode crossings of the $\mathrm{TM_{010}}$ cavity mode with other modes, mostly $\mathrm{TE}$ modes. These mode crossings were predicted via simulation and verified on-site via wide network analyzer scans. There are two major challenges associated with mode crossings. The first is that the form factor diminishes as the cavity mode draws near. The second is that tracking the cavity mode becomes difficult as the other mode appears in the transmission and reflection scans. These issues were circumvented by maneuvering the rods in an anti-symmetric configuration; in other words, moving the rods in opposite direction simultaneously. Moving rods anti-symmetrically shifted several weakly tuning modes, and therefore mode crossings, on the order of a few MHz. This configuration provided form factors around 0.35, sufficient for axion data-acquisition in the previously inaccessible frequency range. Data were acquired in three mode crossing regions using this technique after the initial axion search. An example of anti-symmetric motion as compared to standard symmetric motion can be seen in Fig.~\ref{fig:rod_motion}. The five remaining mode crossings either proved intractable to the changed rod configuration or were too wide to be realistically filled in with this approach.  These can be seen in Table~\ref{tab:regions}.

\begin{table}
    \centering
    \begin{tabular}{|c|c|l|}
    \hline
    Mode Crossing Frequency (MHz) & Width (MHz) \\
    \hline
      704.659 & 0.350  \\
    715.064 & 0.140  \\
    717.025 & 0.140 \\
    726.624 & 0.701  \\
    753.844 & 12.682 \\
      \hline
    \end{tabular}
    \caption{Mode crossing locations where an exclusion limit could not be set.}
    \label{tab:regions}
\end{table}

\section{Rescan Procedure}
\label{sec:rescan_procedure}

A well-defined rescan protocol is critical to the success of any resonant haloscope experiment insofar as it minimizes the chances of missing a potential axion signal. Conditions change throughout the course of the run, and decisions must be made so that a thorough search is conducted regardless. ADMX Run 1B proceeded as follows. The full run frequency range of approximately 125 MHz was scanned in 10-MHz nibbles. This approach enabled us to perform rescans under operating conditions that were similar to the initial scan and kept rescans at a manageable size. After data were acquired from the first pass, the rods were moved in the opposite direction to perform the first rescan. During a rescan, rod motion was slowed and digitization turned on when passing over frequencies flagged as candidates. The following criteria were used to define rescan regions in a grand spectrum:

\begin{enumerate}
     \item The power at that frequency is in excess of 3$\sigma$.
    \item The expected signal-to-noise for a DFSZ axion at that frequency is too low.
    \item Limits set at that frequency do not meet DFSZ sensitivity requirement. In other words, the measured power plus some fraction of sigma (called the candidate threshold power) exceeds the DFSZ axion power.
\end{enumerate}
Regions with SNR less than 2.4 were considered to have insufficient data, triggering a rescan. This particular value was selected because it resulted in a reasonable amount of candidates after a first pass through a nibble. A rescan was also triggered if a candidate's power exceeded that of a DFSZ axion by 0.5$\sigma$.
A persistent candidate is one which does not average to zero with increasing rescans. A true axion signal would not only fulfill this requirement, but its power would maximize on the cavity $\mathrm{TM_{010}}$ mode, with the power scaling as $B^2$. Thus, should a persistent signal maximize on-resonance, the next step in confirming an axion signal would be to switch to the $\mathrm{TM_{011}}$ mode or change the magnetic field and verify the power scaling. The three persistent candidates found in Run 1B are shown in Table~\ref{tab:candidates}. Of the three, one was verified as an initially blinded hardware synthetic, and the other two maximized off-resonance with the $\mathrm{TM_{010}}$ mode, and therefore could not be axions. These other signals were not confirmed to exist independently in the ambient lab setting, although this is perhaps not surprising as the ADMX receiver chain is more sensitive than any ordinary lab equipment. The hardware synthetic maximized on-resonance, but before a magnet ramp could be performed the injection team notified the collaboration that it was in fact a synthetic signal.

\begin{table}
    \centering
    \renewcommand{\arraystretch}{1.5} 
    \begin{tabular}{@{}lcr@{}}
    \hline
    \toprule[0.1ex]
    \hline
        Frequency (MHz) & Notes & Power (DFSZ) \\
        \toprule[0.1ex]
        \hline
        780.255 & Maximized off-resonance & 1.49\\
        730.195 & Synthetic blind & 1.51\\
        686.310 & Maximized off-resonance & 2.36\\
        \hline
        \bottomrule[0.1ex]
        \hline
    \end{tabular}
    \caption{Candidates that persisted past rescan. The signal power of the candidate is shown on the right-hand side, in units of DFSZ signal power.}
    \label{tab:candidates}
\end{table}
\section{Run 1B Limit}
\label{sec:limit}
At the end of all data-taking for Run 1B, the final limit was computed. An RF bin containing an axion signal, scanned multiple times, would result in a Gaussian distribution centered about some mean, $\mu=g_{\gamma}^2\eta$, where $\eta$ is the SNR for the given measurement.  An RF bin containing no axion signal, scanned multiple times, would result in a Gaussian distribution centered about a mean $\mu=0$. A limit was set by computing $\mu$ for a given RF frequency bin that gave a 90\% confidence limit that our measurement did not contain an axion.

It is not obvious from this procedure how to convert a negative power to a limit on $g_{\gamma}$. Thus, in determining the value of $\mu$ that gives the desired confidence level, the cumulative distribution function for a truncated normal distribution was used. This gave a confidence level that covered only physical values of $g_{\gamma}$~\cite{Feldman1998}.

\begin{figure*}
    \centering
    \includegraphics[width=0.8\textwidth]{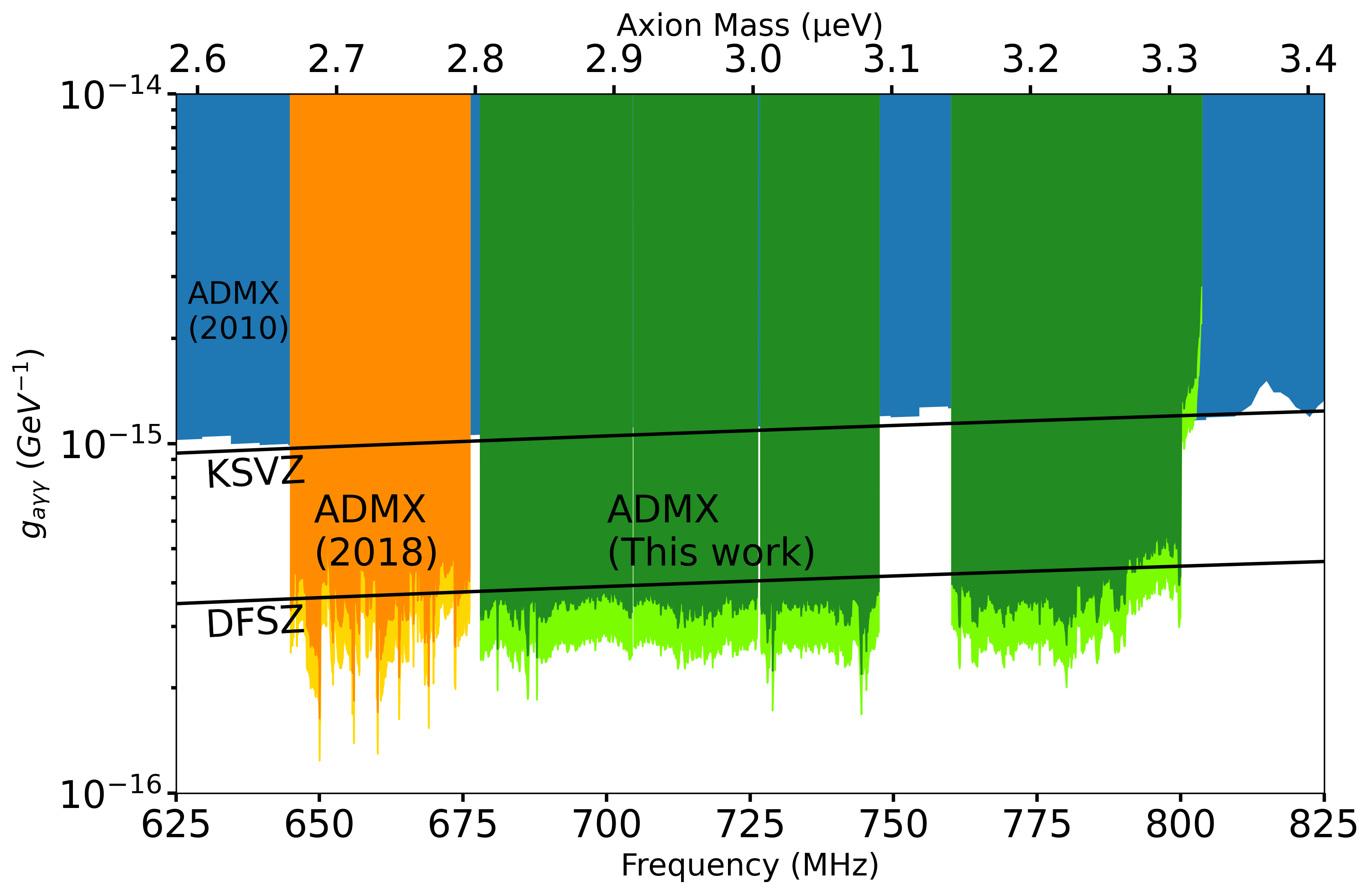}
    \caption{Exclusion plot for Run 1B, shown in green. Dark green represents the region excluded using a standard Maxwell-Boltzmann filter, whereas light green represents the region excluded by an N-body filter~\cite{0004-637X-845-2-121}. }
    \label{fig:exclusion}
\end{figure*}

Because this technique also results in a jagged, 300-Hz bin-wide limit, the following  approach is used to smooth the result to produce an exclusion plot. A small number of bins (200, representing the number of bins in one plot pixel) were combined into a single limit as follows. For each bin, a normal distribution was generated using the measured power as the mean, and the measured uncertainty as the standard deviation. This distribution was then randomly sampled 100 times for each bin. When this resulted in a negative value, it was clipped to zero. The full list of randomly sampled values is then sorted, and the 90\% confidence limit was determined to be the generated power that was 90\% of the way to the top of the sorted list.  

With Run 1B, ADMX was able to exclude the regions shown in green in Fig.~\ref{fig:exclusion}. Dark green shows the region excluded by using the standard Maxwell-Boltzmann filter, whereas light green shows the region excluded by using an N-body filter (see Ref.~\cite{0004-637X-845-2-121}). The Maxwell-Boltzmann exclusion limit used a local dark matter density of 0.45 $\mathrm{GeV/{cm^3}}$, whereas the N-body filter used a local dark matter density of 0.63 $\mathrm{GeV/{cm}^3}$. Regions where there are gaps in the data are due to mode crossings. The frequency range for QCD axions as 100\% dark matter 680-790 MHz was excluded at the 90\% confidence limit, except for the few regions where there were mode crossings. The total mass range covered in Run 1B is larger by a factor of four over the previous Run 1A~\cite{PhysRevLett.120.151301}.

\section{Conclusion}
In conclusion, the ADMX collaboration did not observe any persistent candidates which fulfilled the requirements for an axion signal throughout the course of Run 1B. This implies the 90\% confidence limit exclusion of DFSZ axions for 100\% dark matter density over the frequency range 680-790 MHz (2.81--3.31 $\si\micro$eV), omitting the five regions with mode crossings. Notably, the ADMX collaboration is the only collaboration to have achieved sensitivity to DFSZ axions in this frequency range, and have refined their approach in covering a wider portion of the expected DFSZ axion frequency space than ever before.

\section{Acknowledgements}
This work was supported by the U.S. Department of Energy through Grants No. DE-SC0009800, No. DESC0009723, No. DE-SC0010296, No. DE-SC0010280, No. DE-SC0011665, No. DEFG02-97ER41029, No. DEFG02-96ER40956, No. DEAC52-07NA27344, No. DEC03-76SF00098, and No. DE-SC0017987. Fermilab is a U.S. Department of Energy, Office of Science, HEP User Facility. Fermilab is managed by Fermi Research Alliance, LLC (FRA), acting under Contract No. DE-AC02-07CH11359. Additional support was provided by the Heising-Simons Foundation and by the Lawrence Livermore National Laboratory and Pacific Northwest National Laboratory LDRD offices. LLNL Release No. LLNL-JRNL-813347.
\newpage

\bibliographystyle{apsrev4-1}
\raggedright
\bibliography{references}
\end{document}